%% file: apstemplate.tex
\documentclass[aps,prd,reprint,superscriptaddress,amsmath, amssymb]{revtex4-1}

\usepackage{graphicx}
\usepackage{color}
\usepackage{outlines}

\begin{document}

\preprint{APS/123-QED}

\title{Neutron-Antineutron Oscillation Search using a 0.37 Megaton$\cdot$Year Exposure of Super-Kamiokande}

\include{Authors-20200529}
\date{\today}

\begin{abstract}

As a baryon number violating process with $\Delta B=2$, neutron-antineutron oscillation ($n\to\bar n$) provides a unique 
test of baryon number conservation.
We have performed a search for $n\to\bar n$ oscillation with bound neutrons in Super-Kamiokande, with the full data set from its first four 
run periods, representing an exposure of 0.37~Mton-years. 
The search used a multivariate analysis trained on simulated $n\to\bar n$ events and atmospheric neutrino backgrounds and  
resulted in 11 candidate events with an expected background of 9.3 events.  
In the absence of statistically significant excess, we derived a lower limit on $\bar n$ appearance lifetime in $^{16}$O nuclei 
of $3.6\times{10}^{32}$ years and on the neutron-antineutron oscillation time of $\tau_{n\to\bar n} > 4.7\times10^{8}$~s at 90\% C.L..
\begin{description}
     \item[Key words]Neutron-antineutron oscillation; Super-Kamiokande; Baryon number violation
      \item[DOI]
   \end{description}
\end{abstract}

\maketitle

\section{Introduction}

The present baryon asymmetry of the universe provides indirect evidence for baryon number violating (BNV) 
processes~\cite{Sakharov:1967dj}, which cannot be sufficiently explained by mechanisms 
within the Standard Model (SM)~\cite{Fukugita:1986hr}. 
Searches for BNV processes probe physics beyond the reach of the SM 
can be classified based on the baryon number violation ($\Delta B$) involved.
Processes with $\Delta B = 1$ are tightly constrained by null observations from proton decay searches, 
and processes with $\Delta B = 3$ are expected to conflict with nucleosynthesis scenarios~\cite{Phillips:2014fgb}.
The Standard Model allows for non-perturbative processes involving sphalerons
that would wash out any baryon number asymmetry from processes that conserve $B-L$, where $L$ is lepton number, before the electroweak phase transition~\cite{Kuzmin:1985mm}. 
Therefore, as a BNV process violating both $B$ and $B-L$, neutron-antineutron oscillation provides a unique probe of baryon number violation and essential insight into the baryon asymmetry and baryogenesis.

Since the 1970's several models predicting $n-\bar{n}$ oscillations have been proposed, 
including those employing an SU(2)$_L\times$SU(2)$_R\times$SU(4)$_c$ gauge group to generate a baryon asymmetry~\cite{Babu_2006, Babu:2013yca}
and others that propagate SM fields into extra space-time dimensions~\cite{Nussinov:2001rb}.
The predicted oscillation times vary from $10^9$ s~\cite{Nussinov:2001rb} to $5\times10^{10}$ s~\cite{Babu:2013yca} 
and correspond to energy scales of $10^2\sim10^3$ TeV, well above the scale that can currently be probed by accelerators. 

The probability of a free neutron oscillating to an antineutron can be parameterized as a simple $2\times2$ Hamiltonian and can be written as 
\begin{equation}
	P_{n\to\bar n}(t)=\frac{\delta m^2}{\Delta E^2+\delta m^2}\sin^2(\sqrt{\Delta E^2+\delta m^2}t),
	\label{prob}
\end{equation}
\noindent where $\Delta E$ is the energy difference between the neutron and antineutron and $\delta m=1/\tau_{n-\bar n}$, where $\tau$ is the neutron-antineutron oscillation time.
In the case of degenerate neutron and antineutron energies, Equation.~(\ref{prob}) has a simplified form,
\begin{equation}
	P_{n\to\bar n}(t)\approx(\delta mt)^2=\left(\frac{t}{\tau_{n-\bar n}}\right)^2.
\end{equation}
\noindent For bound neutrons in nuclei, the probability can be written as~\cite{Friedman:2008es} 
\begin{equation}
	P_{\text{nuc}}(n\to\bar n)=\frac{1}{T_{\text{nuc}}}\approx \frac{1}{R\tau_{n-\bar n}^2},
	\label{eq:tranlim}
\end{equation}
\noindent
where $T_{\text{nuc}}$ is the observed neutron lifetime in neutron-antineutron oscillation, and $R$ is the so-called nuclear suppression factor that 
accounts for the suppression of oscillations due to differences in the nuclear potentials of neutrons and antineutrons.
Theoretical calculations of $R$ using effective field theories vary~\cite{Haidenbauer_2020, Oosterhof_2019}, 
but in the following, we adopt $R=0.517\times10^{23}\text{ s}^{-1}$ for $^{16}$O based calculations by Friedman $et. al.$~\cite{Friedman:2008es}.

Experimental searches for $n-\bar n$ oscillation rely on observing particles (mostly pions) produced when a neutron oscillates into an antineutron and annihilates with a nearby nucleon.
There have been a number of $n-\bar n$ searches using either free neutrons~\cite{BaldoCeolin:1994jz} or bound neutrons~\cite{Abe:2011ky, Aharmim:2017jna, Chung:2002fx, Takita:1986zm, Jones:1983ij, Jeffrey:2016, Berger:1989gw}, none of which have yielded a positive signal. 
Accordingly, constraints on the $n-\bar n$ oscillation time have been set at $\tau_{n-\bar n}>0.86\times10^8$ s for free neutron oscillation~\cite{BaldoCeolin:1994jz} and at $\tau_{n-\bar n}>2.7\times10^8$ s for bound neutrons~\cite{Abe:2011ky}.

In this paper, we present a search for $n-\bar n$ oscillations using the full data set from the first four running periods of Super-Kamiokande 
and update the result presented in Ref.~\cite{Abe:2011ky} which used data from the first period.
The current analysis includes an updated data set, an updated hadron production model, final state interactions, 
and adopts a  multivariate method to achieve better discrimination between the background and signal processes. 
This paper is organized as follows.
After a short description of the Super-Kamiokande detector in Section~\ref{sec:SK}, we describe the simulation of both the $n-\bar n$ signal and 
atmospheric neutrino background in Section~\ref{sec:sim}.
The selection algorithm and analysis cuts are explained in Section~\ref{sec:reco}, followed by discussion of 
systematic uncertainties in Section~\ref{sec:sys}.
Analysis results and concluding remarks are presented in Sections~\ref{sec:result} and \ref{sec:conclusion}, respectively.

\section{The Super-Kamiokande experiment}
\label{sec:SK}

Super-Kamiokande (SK) is a cylindrical 50~kiloton water Cherenkov detector located in Kamioka, Japan, that is 
shielded by a 2,700 meter water-equivalent rock overburden~\cite{Fukuda:2002uc}.
The detector consists of an outer detector (OD) instrumented with 1885 outward-facing 8-inch PMTs mounted 2~m from the detector's outer wall 
on a structure that optically separates it from the inner detector (ID). 
This structure also supports the 11,129 inward-facing 20-inch PMTs that form the ID and view its 32~kton target volume.
The OD is primarily used as a veto for charged particles entering from outside the detector or identifying particles 
that exit the ID, and the ID itself is used reconstruct the energies, vertexes, and particle types of most interest to the present work.
 
The experiment started data taking in 1996 and underwent four data-taking phases since then labeled as SK-I, II, III, and IV.
The SK-I period ran from 1996 until the detector underwent maintenance in 2001.
During that period, an accident destroyed more than half of the SK PMTs, reducing the photocathode coverage 
from $\sim 40\%$ to $\sim 19\%$ for the SK-II period in 2002-2005.
After replacing the missing PMTs in 2005, the detector restarted operations as SK-III in 2006-2008.
Following upgrades of the front-end electronics and water purification system, the SK-IV period ran from 
2008 until May of 2018, when the data taking was paused and the detector tank was opened for further upgrades.
The analysis in this work uses the full data set from the SK-I-IV periods.
Details of the detector and its calibration can be found in~\cite{Abe_2014}. 
 
\section{Simulation}
\label{sec:sim}

Following the oscillation of a neutron into an antineutron, the subsequent annihilation of the antineutron with a nucleon in the oxygen nucleus is expected to produce many visible particles, most of which are pions. 
The simulation of this signal is broken into stages: oscillation, hadronization, final state interactions of particles before exiting the nucleus, and finally propagation and subsequent reinteraction of those particles with detector media.
During the first stage, the position of the oscillated neutron within the nucleus is determined using the standard Woods-Saxon distribution~\cite{Woods:1954zz,Friedman:2008es} with a Fermi momentum simulation based on the spectral function measured in~\cite{Nakamura:1976mb}.
The effect of nuclear binding energy is taken into account by subtracting it from the nucleon masses when calculating the annihilation products, using 39.0 MeV for s-state and 15.5 MeV for p-state nucleons respectively.
Thereafter the oscillated antineutron is assumed to have an equal probability of annihilating with any remaining nucleons.

Modeling of the $\bar nn$ or $\bar np$ annihilation products is done based on available accelerator data.
Due to a lack of antineutron scattering data, the hadronization simulation uses results from antiproton scattering experiments instead.
Assuming isospin symmetry, we used data from the $\bar pp$ annihilation experiment Crystal Barrel~\cite{Klempt:2005pp, Amsler:2003bq} 
to simulate the $\bar nn$ annihilation. 
For the $\bar np$ channel, we used the $\bar pn$ annihilation branching ratio measurements from the OBELIX experiment~\cite{Bressani:2003pv} and bubble chamber data~\cite{Bizzarri1984, Pavlopoulos:1977mb, Backenstoss:1983gu} and then flipped the signs of the charged pions to match $\bar np$.
Tables~\ref{tab:chnn} and ~\ref{tab:chnp} show the branching ratios for $\bar nn$ and $\bar np$ adopted in the simulation. 
The branching ratios of kaonic channels are artificially constructed due to lack of experimental data, and the kaonic production for $\bar np$ is less than 1/2 from $\bar nn$, and thus is omitted.
Corresponding uncertainty calculations can be found in Section~\ref{sec:sys}, and the efficiency calculation is explained in Section~\ref{sec:reco}.

\begin{table} 
	\centering
	\caption{Branching ratios ($B_R$), relative uncertainties, and corresponding efficiencies for $\bar nn$ annihilation products.}
   \label{tab:chnn}
   \begin{tabular}{lrrr}
      \hline\hline
		& $B_R$ [\%] & Relat. Uncer. & Efficiency [\%] \\
      \hline
		2$\pi_0$ & 0.1 & 5\% & 3.2\\
		\hline
		3$\pi_0$ & 0.7 & 6\% & 3.6\\
		4$\pi_0$ & 0.3 & 6\% & 4.4\\
		5$\pi_0$ & 1.0 & 4\% & 3.8\\
		7$\pi_0$ & 0.1 & 8\% & 2.1\\
      \hline
		$\pi^+\pi^-$ & 0.3 & 4\% & 4.8\\
		$\pi^+\pi^-\pi_0$ & 1.6 & 15\% & 4.8\\
		$\pi^+\pi^-2\pi_0$ & 13.1 & 15\% & 4.3\\
		$\pi^+\pi^-3\pi_0$ & 11.2 & 15\% & 4.2\\
		$\pi^+\pi^-4\pi_0$ & 3.3 & 14\% & 4.0\\
		$\pi^+\pi^-5\pi_0$ & 1.4 & 15\% & 4.7\\
      \hline
		$2\pi^+2\pi^-$ & 6.0 & 16\% & 4.2\\
		$2\pi^+2\pi^-\pi_0$ & 13.6 & 15\% & 4.5\\
		$2\pi^+2\pi^-2\pi_0$ & 15.7 & 15\% & 4.5\\
		$2\pi^+2\pi^-3\pi_0$ & 0.6 & 33\% & 4.9\\
      \hline
		$3\pi^+3\pi^-$ & 2.2 & 15\% & 3.7\\
		$3\pi^+3\pi^-\pi_0$ & 2.0 & 15\% & 4.1\\
		\hline
		$\rho^0\pi^0$ & 1.8 & 15\% & 4.8\\
		$\rho^{+/-}\pi^{-/+}$ & 3.7 & 15\% & 4.5\\
		$\omega\omega$ & 3.5 & 15\% & 4.5\\
		$\rho^0\omega$ & 2.4 & 15\% & 4.0\\
		$\pi^0\pi^0\omega$ & 2.7 & 15\% & 3.8\\
		$\pi^+\pi^-\omega$ & 7.1 & 15\% & 4.5\\
		$\eta\omega$ & 1.6 & 15\% & 4.6\\
		$\pi^+\pi^-\eta$ & 1.7 & 15\% & 3.8\\
      \hline
		Kaonic channels & 2.3 & 15\% & 4.5\\
      \hline\hline
   \end{tabular}
\end{table}

\begin{table} 
	\centering
	\caption{Branching ratios ($B_R$), relative uncertainties, and corresponding efficiencies for $\bar np$ annihilation products.}
   \label{tab:chnp}
   \begin{tabular}{lrrr}
      \hline\hline
		& $B_R$ [\%] & Relat. Uncer. & Efficiency [\%] \\
      \hline
		$\pi^+\pi_0$ & 0.1 & 32\% & 3.4\\
		$\pi^+2\pi_0$ & 0.7 & 32\% & 3.2\\
		$\pi^+3\pi_0$ & 14.8 & 32\% & 3.5\\
		$\pi^+4\pi_0$  & 1.4 & 32\% & 2.6\\
      \hline
		$2\pi^+\pi^-$ & 2.0 & 10\% & 3.6\\
		$2\pi^+\pi^-\pi_0$ & 17.0 & 10\% & 3.5\\
		$2\pi^+\pi^-2\pi_0$ & 10.8 & 10\% & 3.4\\
		$2\pi^+\pi^-3\pi_0$ & 30.1 & 10\% & 3.8\\
      \hline
		$3\pi^+2\pi^-$ & 5.5 & 10\% & 3.2\\
		$3\pi^+2\pi^-\pi_0$ & 3.2 & 10\% & 3.2\\
		\hline
		$\pi^+\pi^0\omega$ & 2.0 & 32\% & 3.4\\
		$2\pi^+\pi^-\omega$ & 12.4 & 32\% & 3.6\\
      \hline\hline
   \end{tabular}
\end{table}

Hadronization products are mostly pions.
The pion interaction probability within the oxygen nucleus is expected to be large, and 
these so-called final state interactions (FSI) include quasi-elastic scattering (e.g., $\pi+n\to\pi+n$), 
absorption ($\pi^++n\to p$, $\pi^-+p\to n$), charge exchange ($\pi^++n\to p+\pi^0$, $\pi^-+p\to n+\pi^0$), 
and pion production ($\pi^\pm+n\to\pi^\pm+n+\pi^0$)~\cite{dePerio:2011zz}.
For pions above 500 MeV/c, the surrounding nucleons are treated as quasi-free particles, while for lower momentum pions the interaction probabilities
are calculated according to the model of Salcedo and Oset~\cite{Salcedo:1987md} in consideration of the effect of Pauli blocking.
More details can be found in Ref.~\cite{dePerio:2011zz}.

Atmospheric neutrino interactions in water are the dominant background to the search for $n-\bar n$ oscillation at SK.
The theoretical calculation from the HKKM model~\cite{Honda:2006qj, Honda:2011nf} predicts the atmospheric neutrino flux at Kamioka in the energy region from sub-GeV up to several TeV after oscillation.
Using this flux prediction, we simulated atmospheric neutrino interactions, including the outgoing particles and their subsequent interactions with the nuclear medium in water, with NEUT version 5.3.6~\cite{Hayato:2009zz}.
Final state interactions for both the signal and background are simulated with NEUT. 
 
Particles escaping the nucleus are passed to a GEANT3-based~\cite{Brun:1994aa} detector simulation.
The simulation tracks particles through the detector medium, simulating their interactions in water as well as 
the production of secondary particles and the response of the PMTs to Cherenkov radiation. 
Detailed tuning and calibration has been performed to provide a tailored simulation of photon propagation in Super-K~\cite{Abe:2013gga}.
The interaction of hadrons with water is simulated using the GCALOR package~\cite{Zeitnitz:1994bs}, except for 
pions below 500 MeV/c, which are simulated using a model based on NEUT's FSI simulation.
The final background is reweighted to the result of the analysis in~\cite{Abe:2017aap}, 
adjusting its central value to the best fit oscillation and systematic error parameters favored by the Super-K data.

\section{Event Reconstruction and Selection}
\label{sec:reco}

The present analysis uses the full data set from the SK-I through SK-IV periods,  corresponding to 6050.0 live-days.
Events are required to be fully contained (FC), meaning the number of PMTs in the highest charge cluster of outer detector hits 
is less than 10~in SK-I and less than 16~in SK-II-III-IV.
Timing information in each event's hit PMTs in the ID is used to reconstruct an overall vertex from the event, 
from which an iterative search based on the Hough transform~\cite{Hough:1959qva} is performed to identify 
Cherenkov rings. 
Each Cherenkov ring is classified according to its hit pattern and opening angle 
as either ``showering'' (e-like) for particles that create electromagnetic showers 
such as $e$ and $\gamma$ or as ``non-showering'' ($\mu$-like) for particles such as $\mu$ and $\pi^\pm$.
The momentum of each ring is determined by the particle type and the charge among all hit PMTs within a 
70 $^\circ$ cone around the ring with consideration of charge shared between multiple rings.
An additional search for delayed electrons from muon decays is performed from 1.2-20~$\mu$s after the 
primary event trigger. 

This analysis starts with FC events more than 2.0~m from the ID wall, which defines a 22.5~kton fiducial volume.
The reduction efficiency is 92\% for $n\to\bar n$ signal events in fiducial volume. 
This sample is then processed in two stages, first applying simple analysis cuts before applying a 
multivariate technique to extract the signal.
 
\subsection{Analysis Pre-cuts}

Based on the distinct features of $n-\bar n$ and atmospheric neutrino events, 
several preliminary cuts are applied to reduce background rates while maintaining high signal efficiency.
The $n-\bar n$ oscillation signal is expected to have multiple pions, while a large number of atmospheric neutrino 
interactions are elastic scatters with only one Cherenkov ring from the outgoing charged lepton.
Therefore, the number of reconstructed rings is required to be $>$1.
This cut removes $\sim$ 75\% of the background while keeping 89\% of the signal.
Unlike the wide range of energies covered by atmospheric neutrinos, the $n-\bar n$ signal is more kinetically constrained, and thus a set of kinematic cuts are also applied.
Here, the total reconstructed momentum is required to be within [35, 875]~MeV/c, the visible energy in [30, 1830]~MeV, 
and the total reconstructed invariant mass in [80, 1910]~MeV/c$^2$.
After the cut on the number of rings, these kinematic cuts further remove $\sim$ 50\% of the background with a relative 
signal efficiency of 98\%.

\subsection{Multivariate Analysis}

\begin{figure} 
   \centering
  \includegraphics[width=0.90\columnwidth]{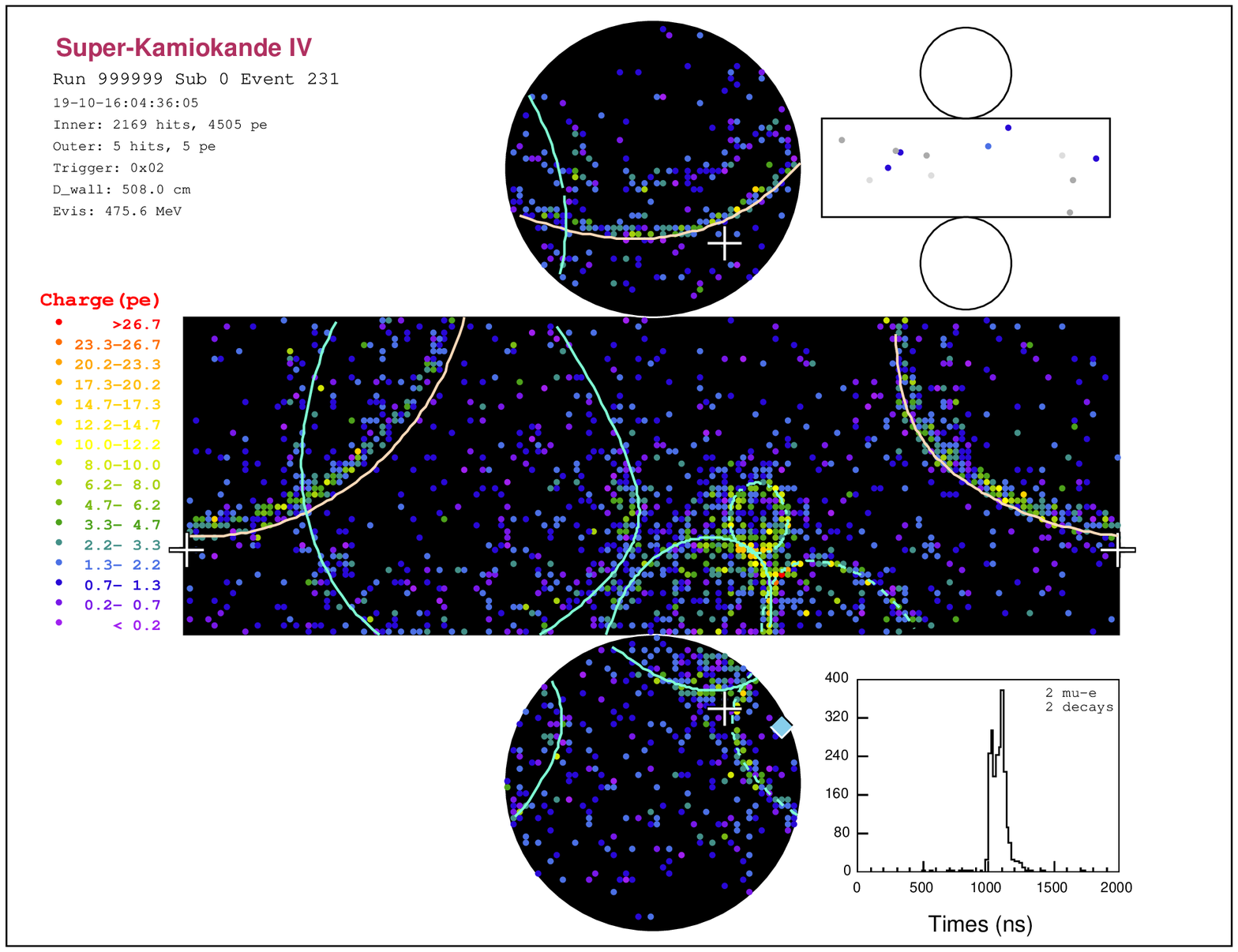}
  \includegraphics[width=0.90\columnwidth]{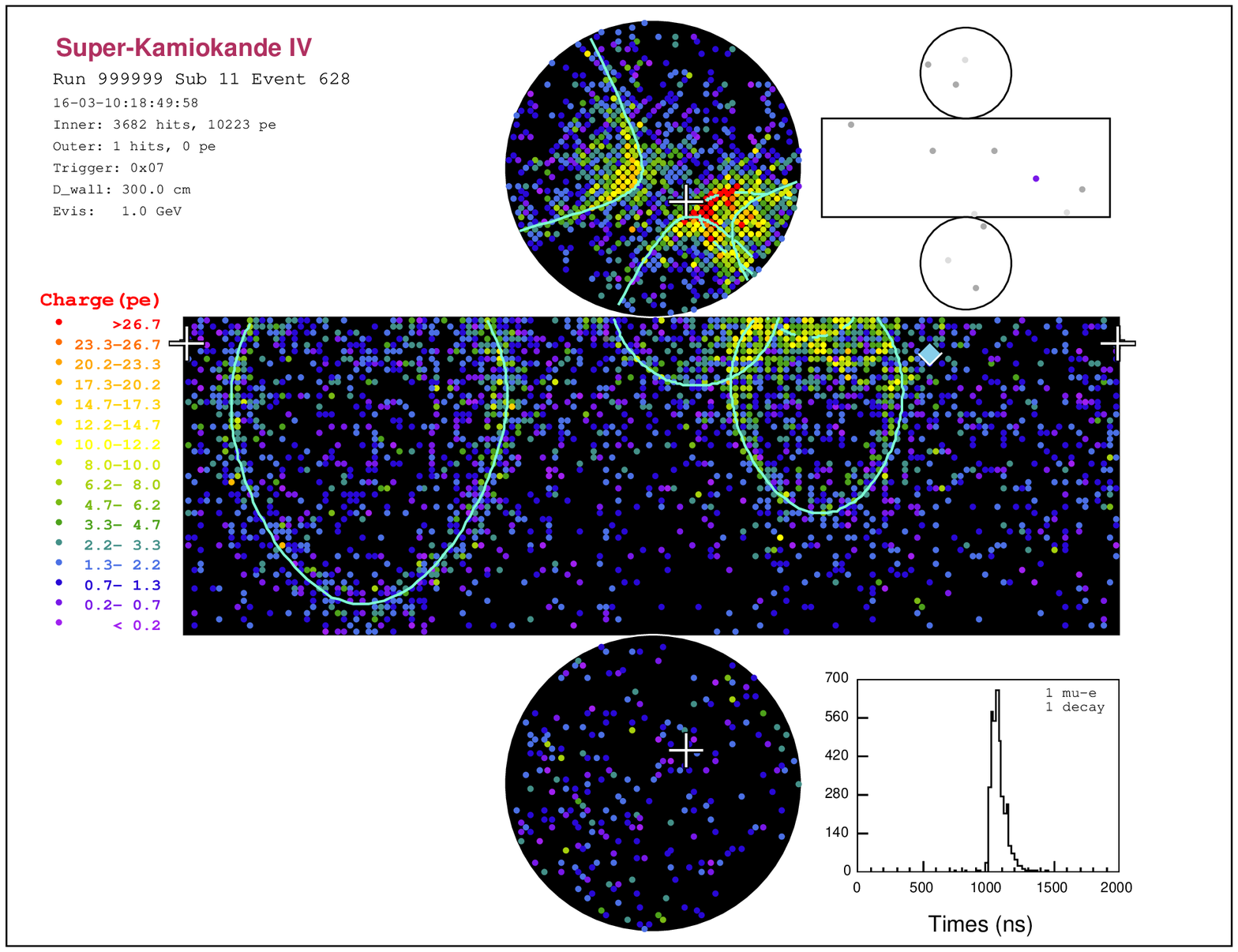}
		\caption{The event display of a simulated $n\to\bar n$ oscillation event (top) and a simulated background event (bottom).
		\\The top figure shows the $\bar np$ annihilation producing 6 pions. In total, there are 13 rings of these pions and their decay products, among which 5 were reconstructed (colored rings).
		The beige ring is a successfully reconstructed $\pi^+$.
		The dashed small ring is a $\pi^-$ mistakenly reconstructed as a electron-like particle, and the dashed large ring is a similarly mis-reconstructed $\pi^+$.
		The two solid cyan rings are 2 $\gamma$'s from the decay of a $\pi^0$.
		\\The bottom figure shows the deep inelastic scattering process of a $\nu_e$. 
		In this event, there are 4 rings reconstructed by the algorithm, all of which are $\gamma$'s.
		}
   \label{fig:eventdisp}
\end{figure}

Event displays of a simulated $n\to\bar n$ signal event and a simulated background event are shown in Fig.~\ref{fig:eventdisp}.
Due to the high ring multiplicity, the performance of ring reconstruction for $n\to\bar n$ signal events is not as satisfactory as typical sub-GeV neutrino events.
To compensate for the limitation of ring reconstruction and to include more discriminant features, we applied a multivariate analysis (MVA) to events passing the pre-cuts.
Compared to a conventional box-cut analysis~\cite{Abe:2011ky}, this analysis significantly enhance the separation between $n\to\bar n$ signal and background.
An estimation using the same MC set shows that the sensitivity of the MVA method is twice that of the box-cut method.

Compared to atmospheric neutrino backgrounds, $\bar nn$ or $\bar np$ annihilation within oxygen are generally expected to 
be more constrained kinematically and have more Cherenkov rings isotropically distributed in the detector.
To exploit these features, we introduced 12 variables into the MVA, 
among which three are conventional kinematic quantities, including the visible energy, total momentum, and total invariant mass.

The remaining nine input variables are as follows.
Since only a fraction of atmospheric neutrinos has sufficient energy to produce multiple charged particles,
signal events are typically expected to have more visible Cherenkov rings.
The number of such rings is used as a variable.
However, the full reconstruction is limited to five rings, as in the case of Fig.~\ref{fig:eventdisp}.
Therefore, an additional variable that counts ring fragments, or potential rings, is also introduced.

The total momentum of an $n-\bar n$ event is limited by the momenta of the interacting nucleons, 
while a background event can carry more momentum from the incident neutrino and is expected to be more forward-going at the energies needed 
to produce multiple particles.
Therefore, this search employs four variables to quantify the isotropy of candidate events.
The energy ring ratio is defined as 
$(E_\text{tot}-E_\text{max})/[E_\text{tot}\cdot(n_\text{ring}-1)]$, 
where $E_\text{max}$ is the energy of the ring with highest energy in an event, 
$E_\text{tot}$ is the total energy of the event, and $n_\text{ring}$ is the number of rings.
For the $n-\bar n$ signal, the annihilation energy is more uniformly distributed among the outgoing pions and 
therefore, the distribution of this variable is expected to have a sharper peak than than of backgrounds.
Signal events are also expected to have higher sphericity than backgrounds, so this analysis 
adopts a sphericity variable~\cite{PhysRevD.1.1416}. 
Fox-Wolfram moments, which are superpositions of spherical harmonics that measure correlations 
between particle momenta (see Ref.~\cite{PhysRevLett.41.1581} for details) are also adopted to describe the correlation between rings.
This analysis employs the first and second order Fox-Wolfram moments, since higher orders 
were found to provide little extra discrimination ability.

\begin{figure*} 
   \centering
  \includegraphics[width=0.24\linewidth]{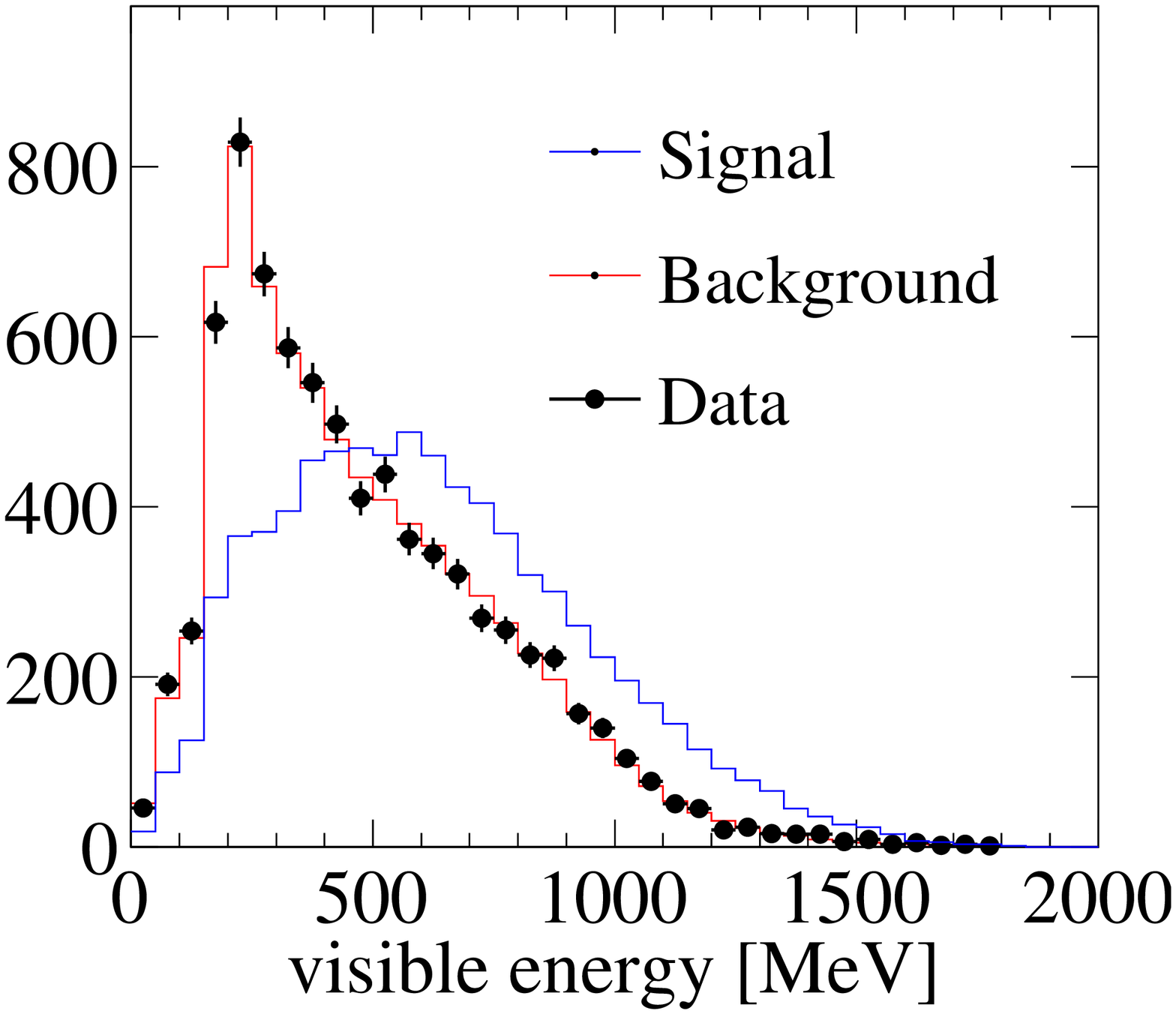}
  \includegraphics[width=0.24\linewidth]{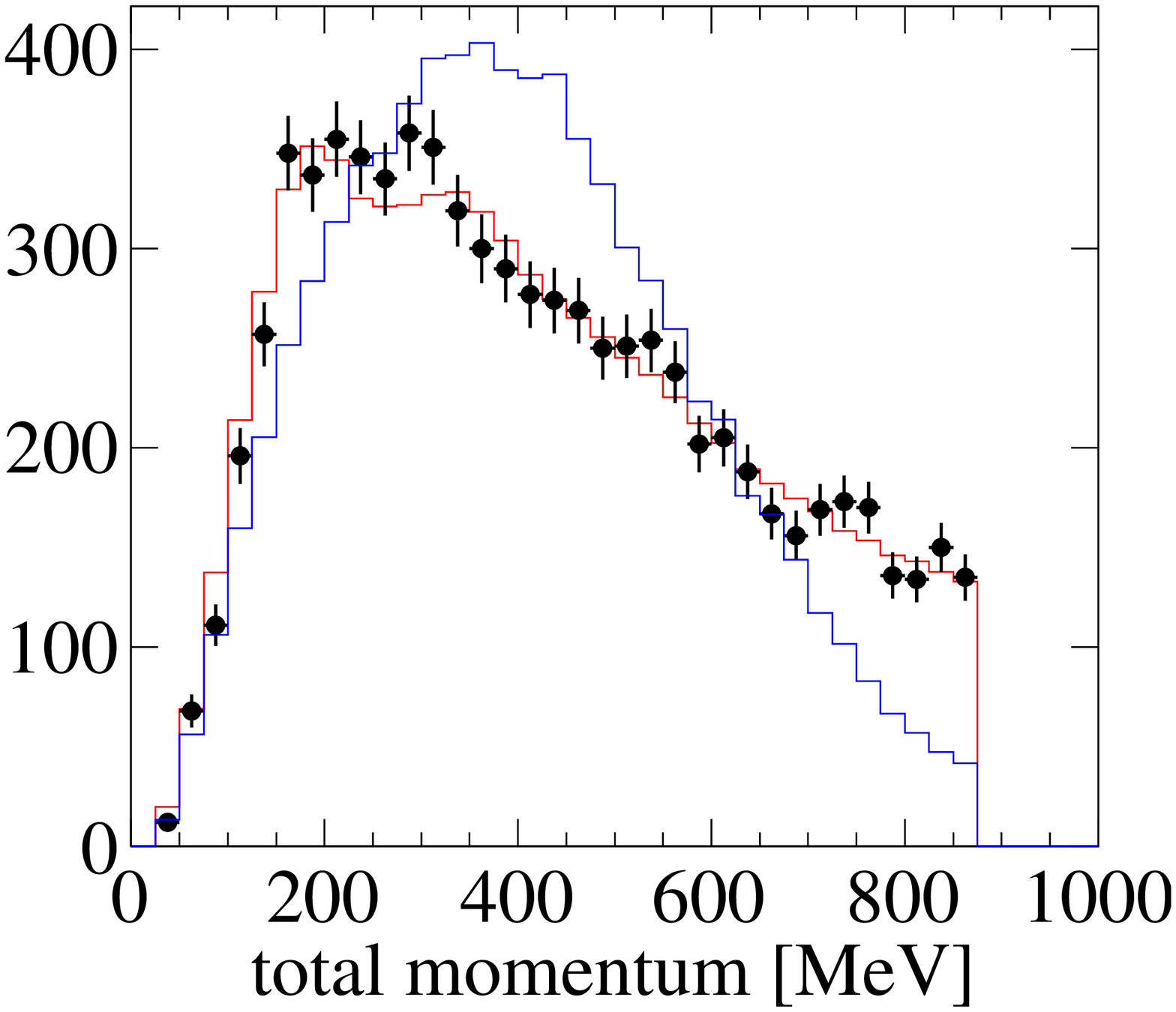}
  \includegraphics[width=0.24\linewidth]{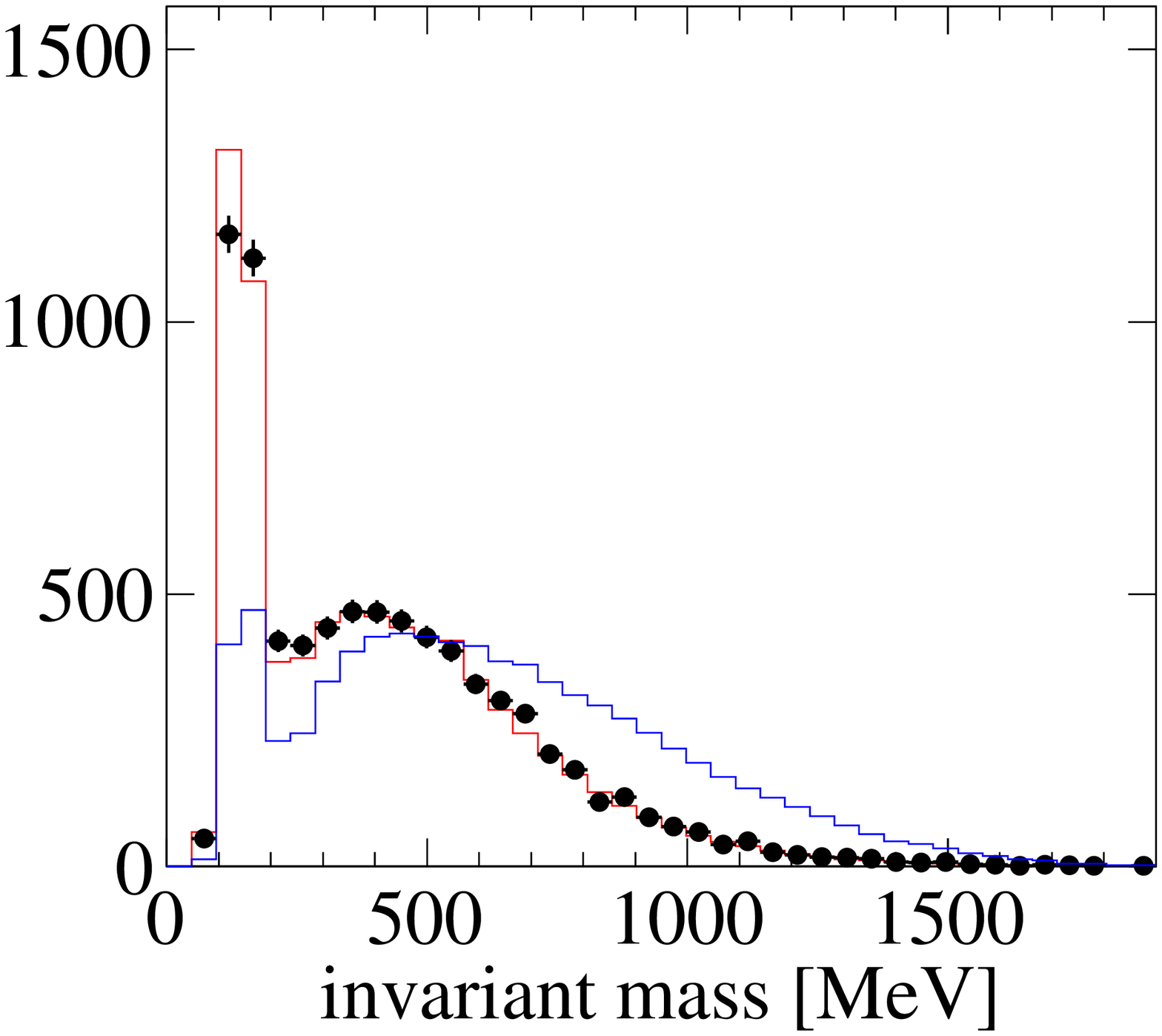}
  \includegraphics[width=0.24\linewidth]{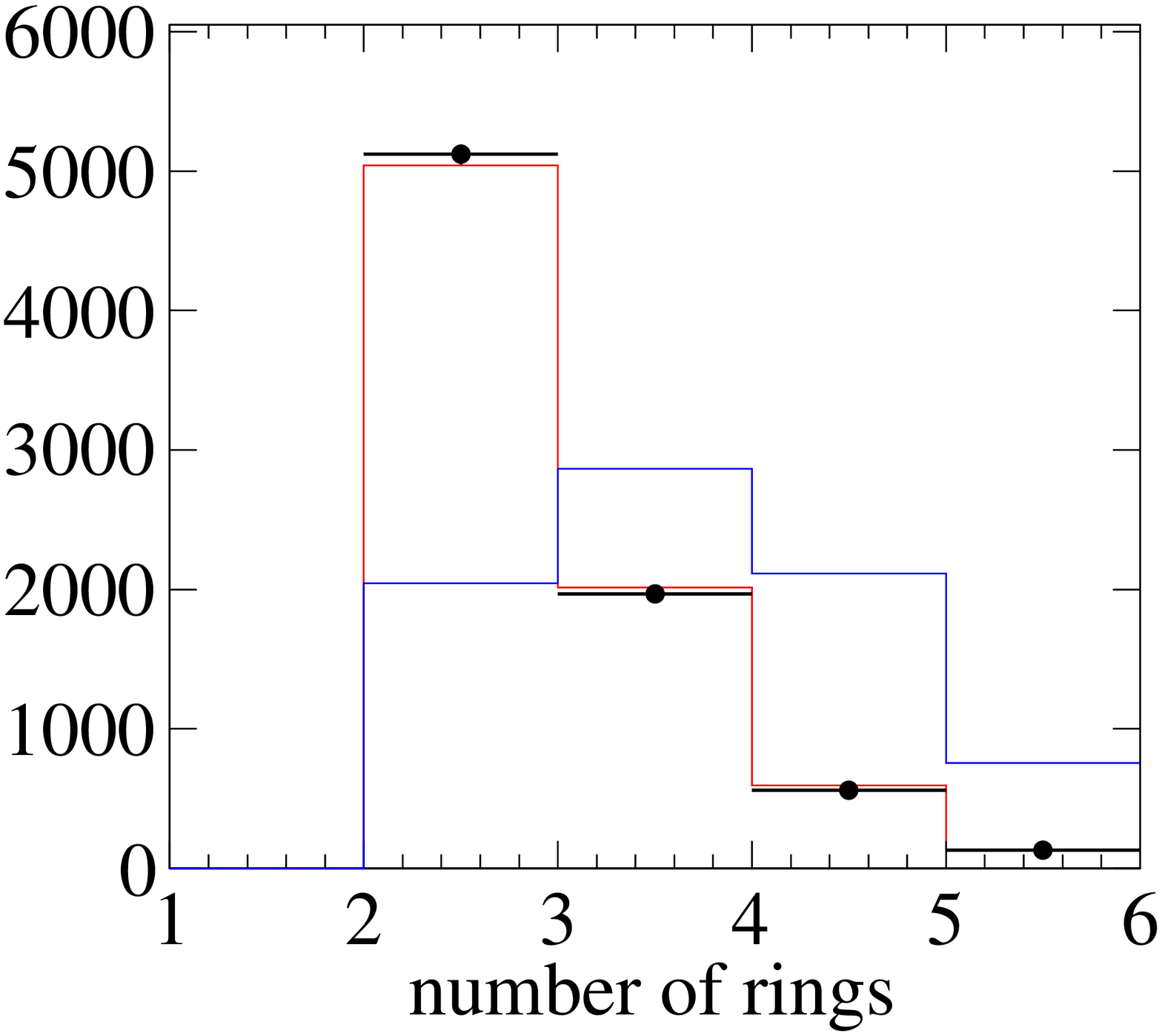}
\\\includegraphics[width=0.24\linewidth]{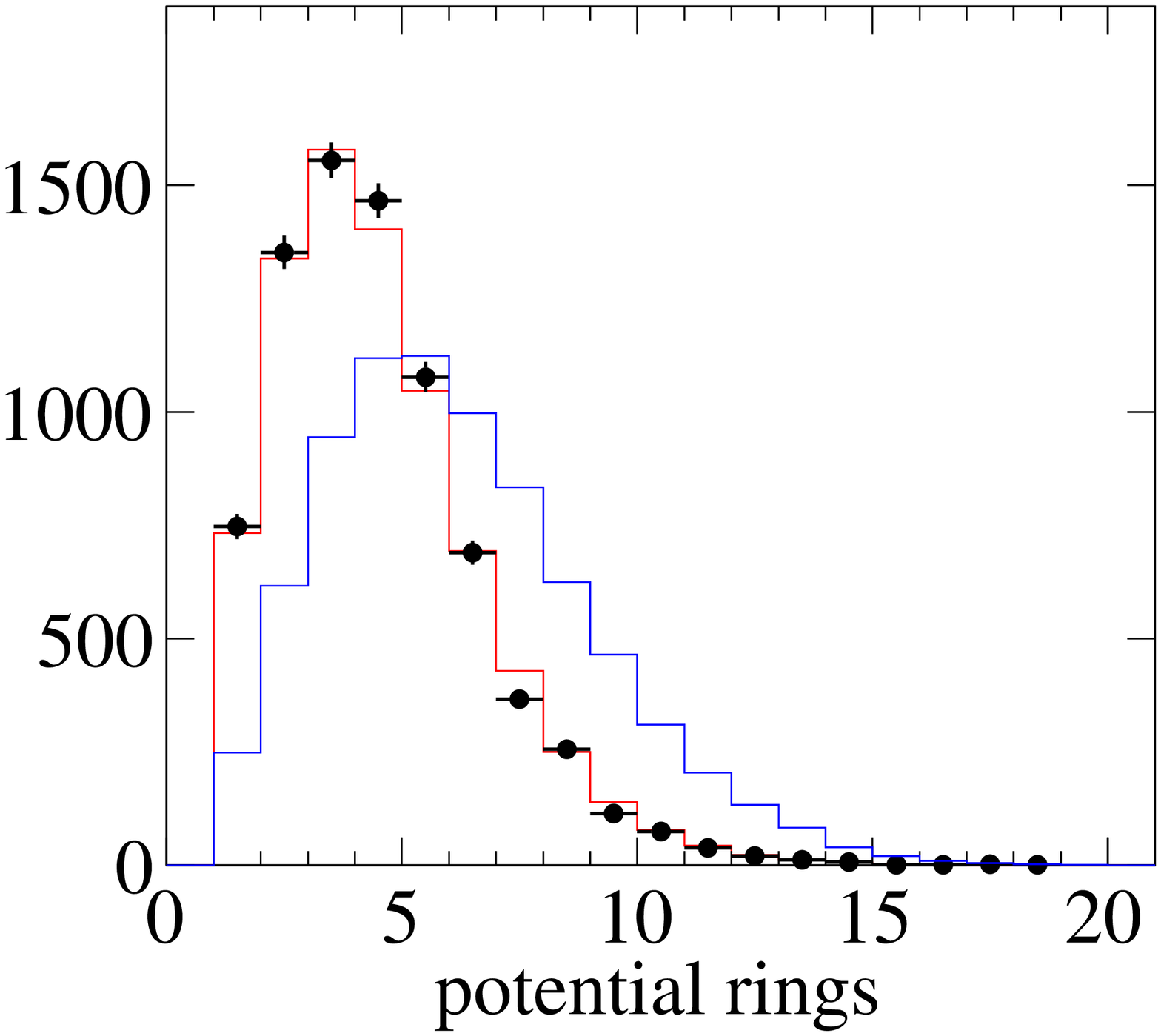}
  \includegraphics[width=0.24\linewidth]{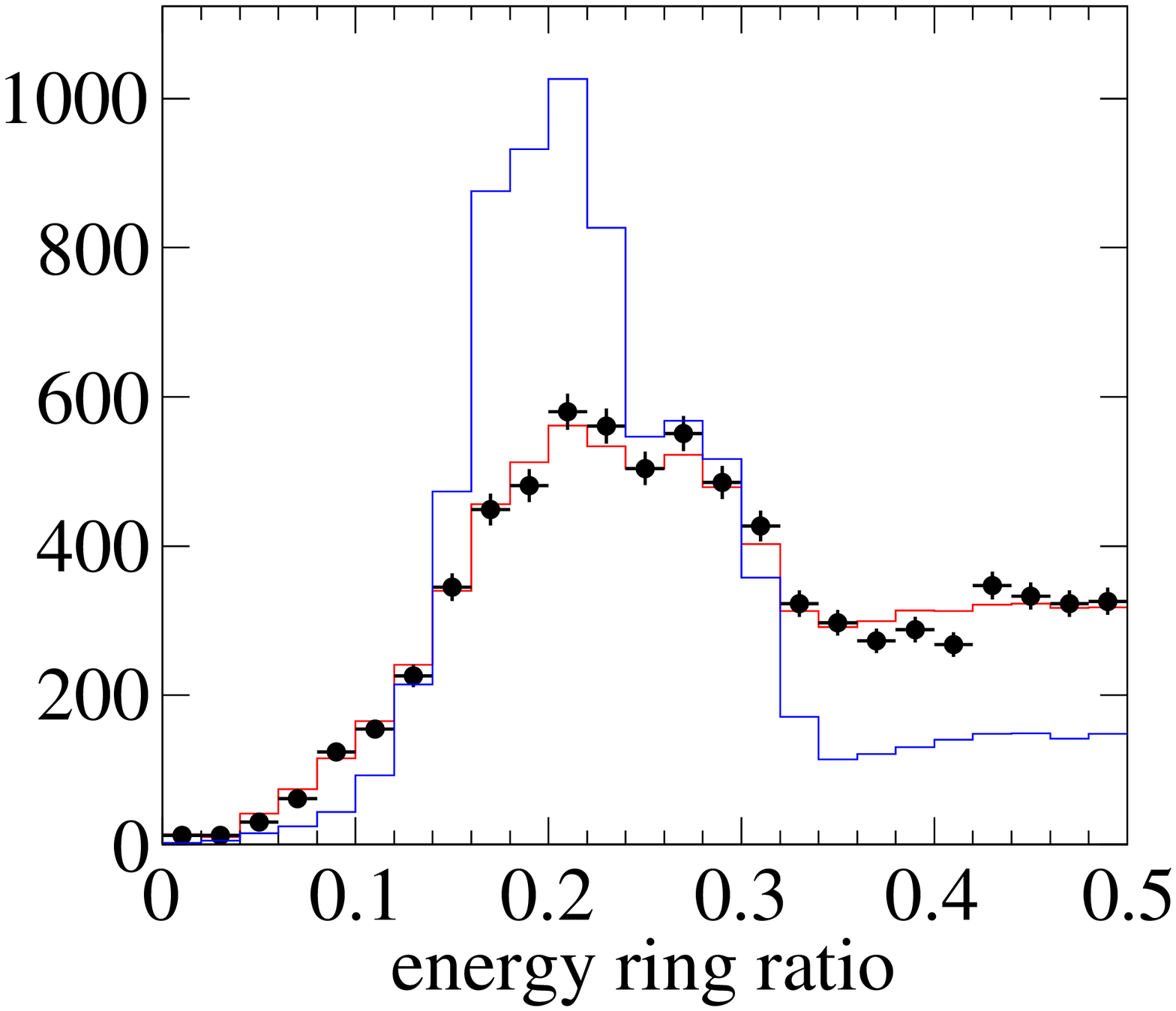}
  \includegraphics[width=0.24\linewidth]{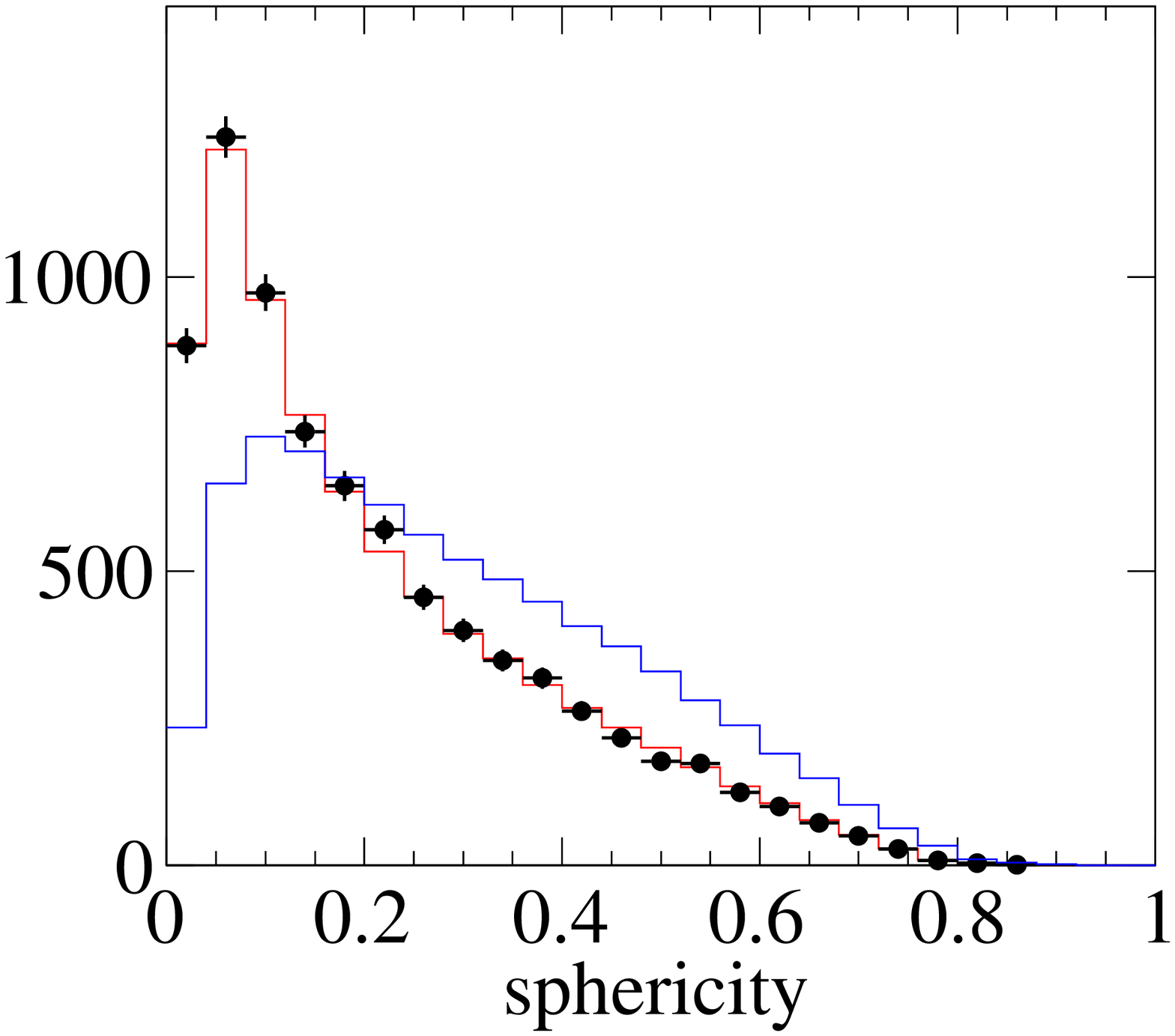}
  \includegraphics[width=0.24\linewidth]{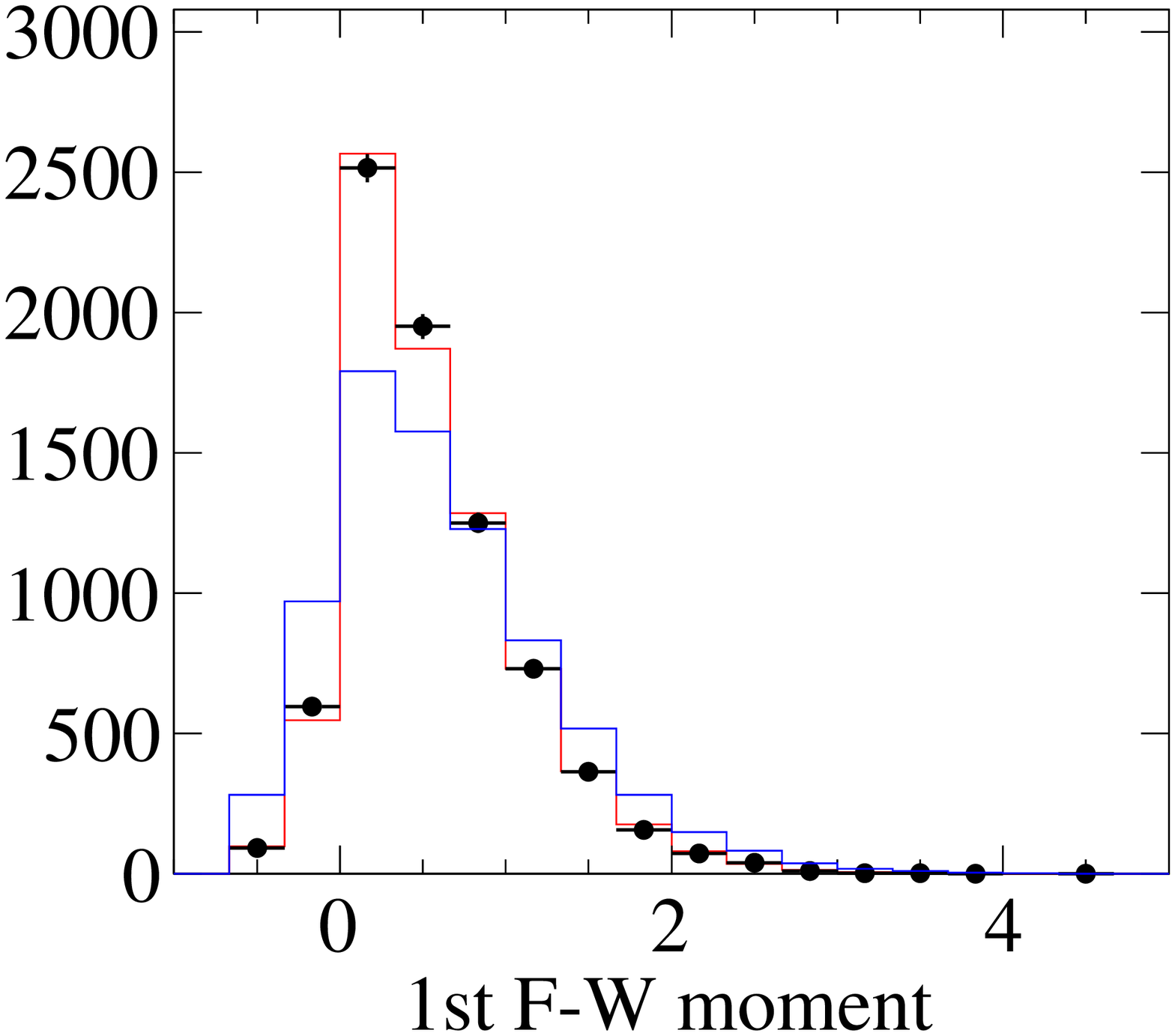}
\\\includegraphics[width=0.24\linewidth]{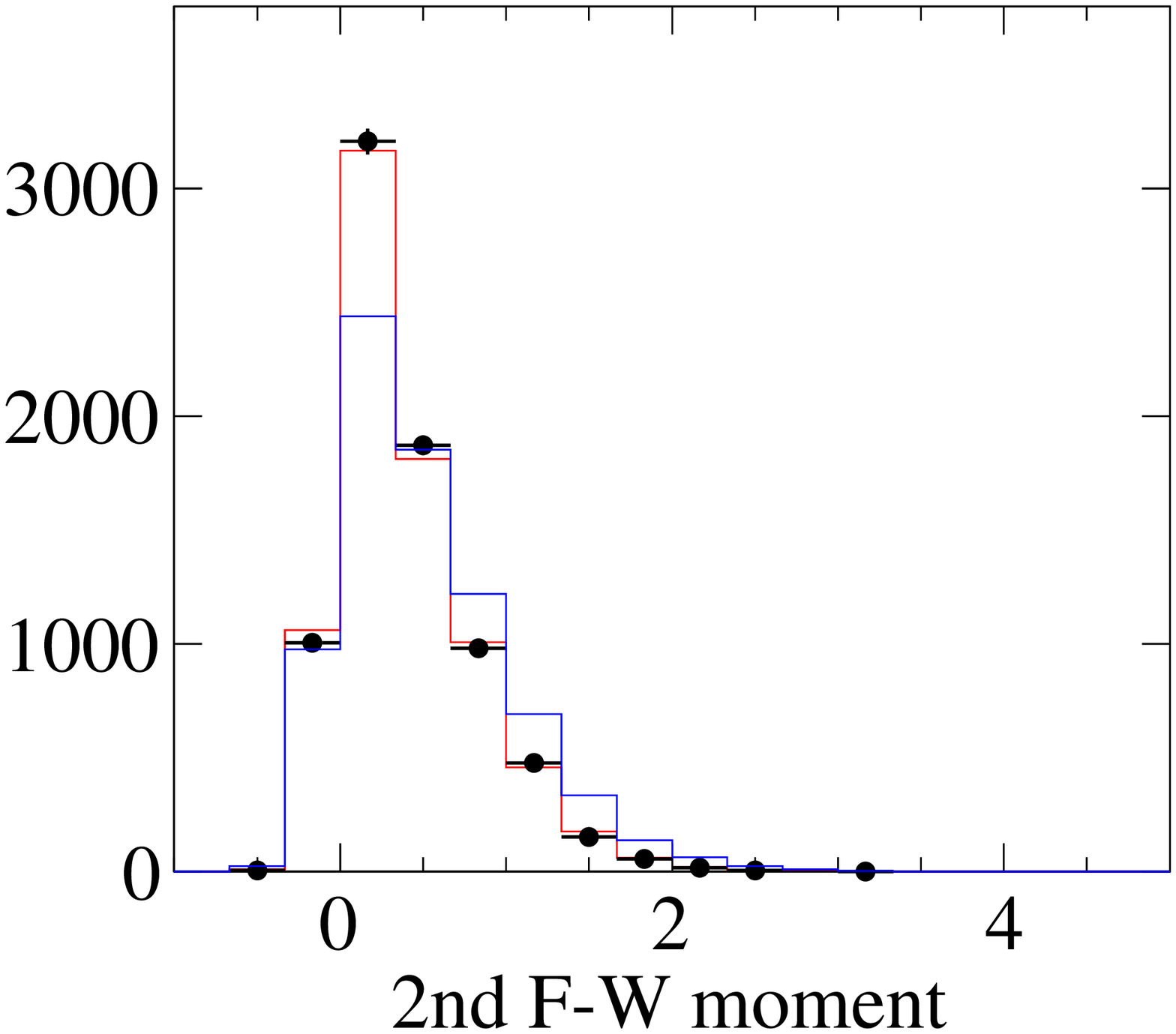}
  \includegraphics[width=0.24\linewidth]{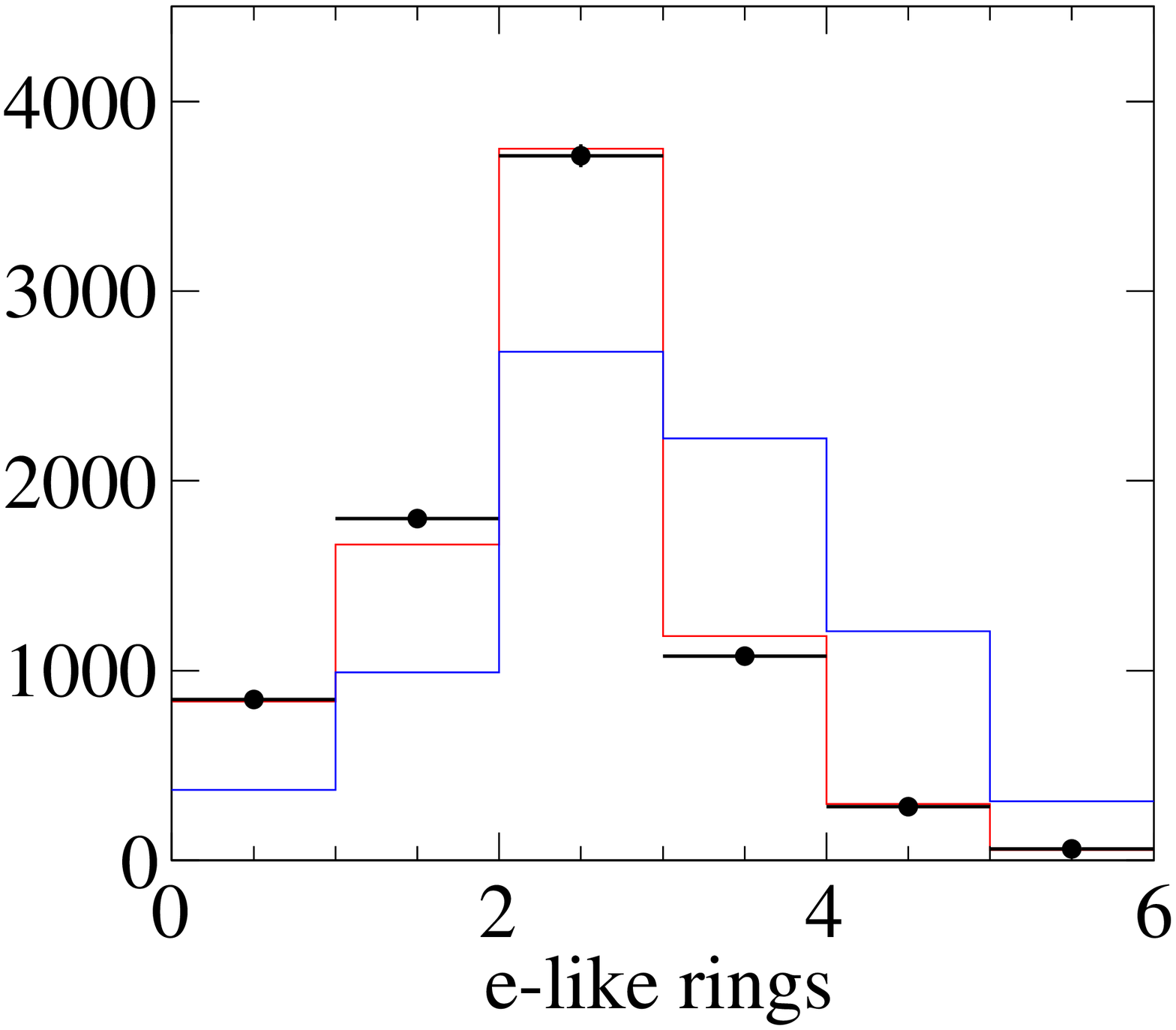}
  \includegraphics[width=0.24\linewidth]{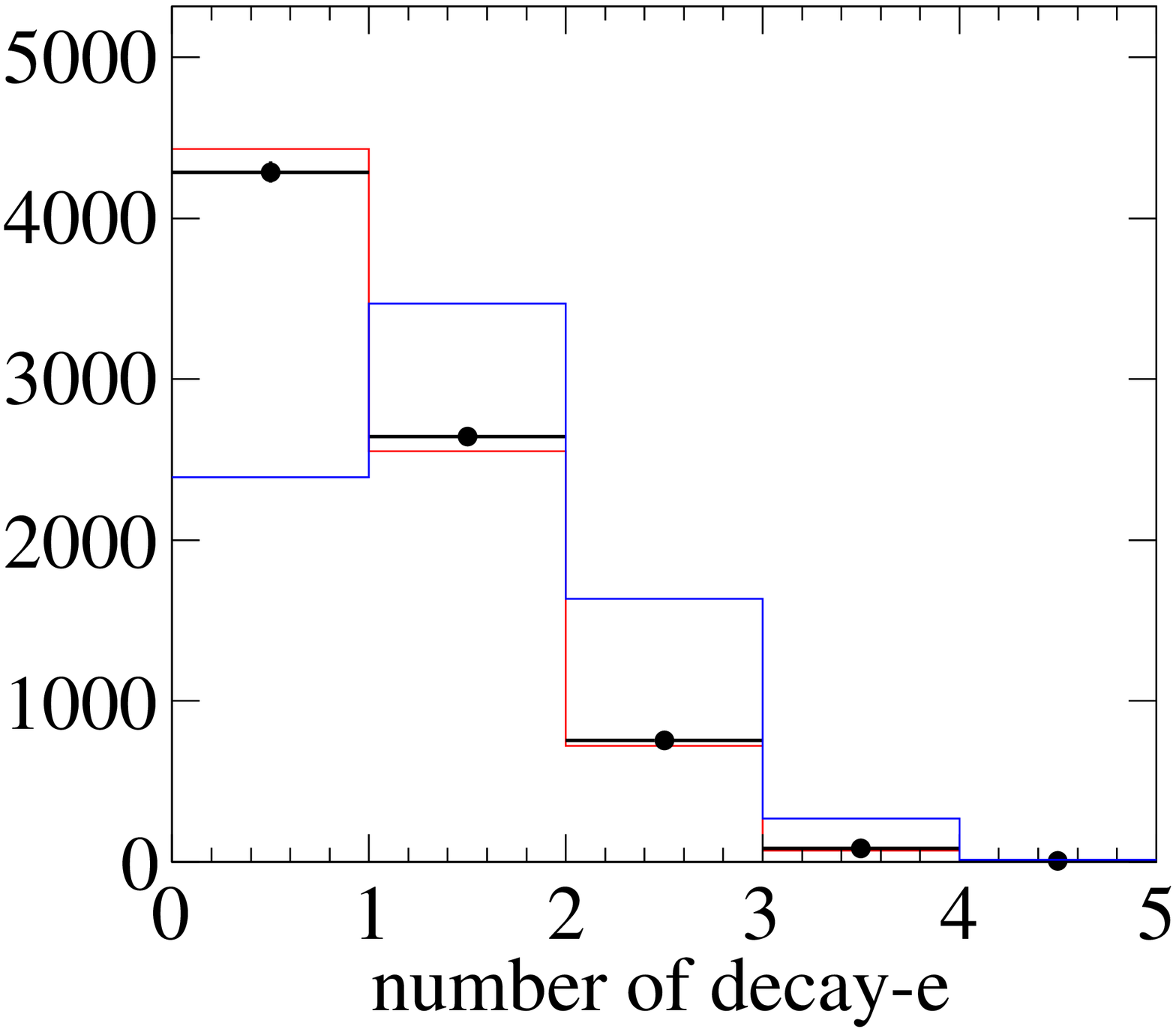}
  \includegraphics[width=0.24\linewidth]{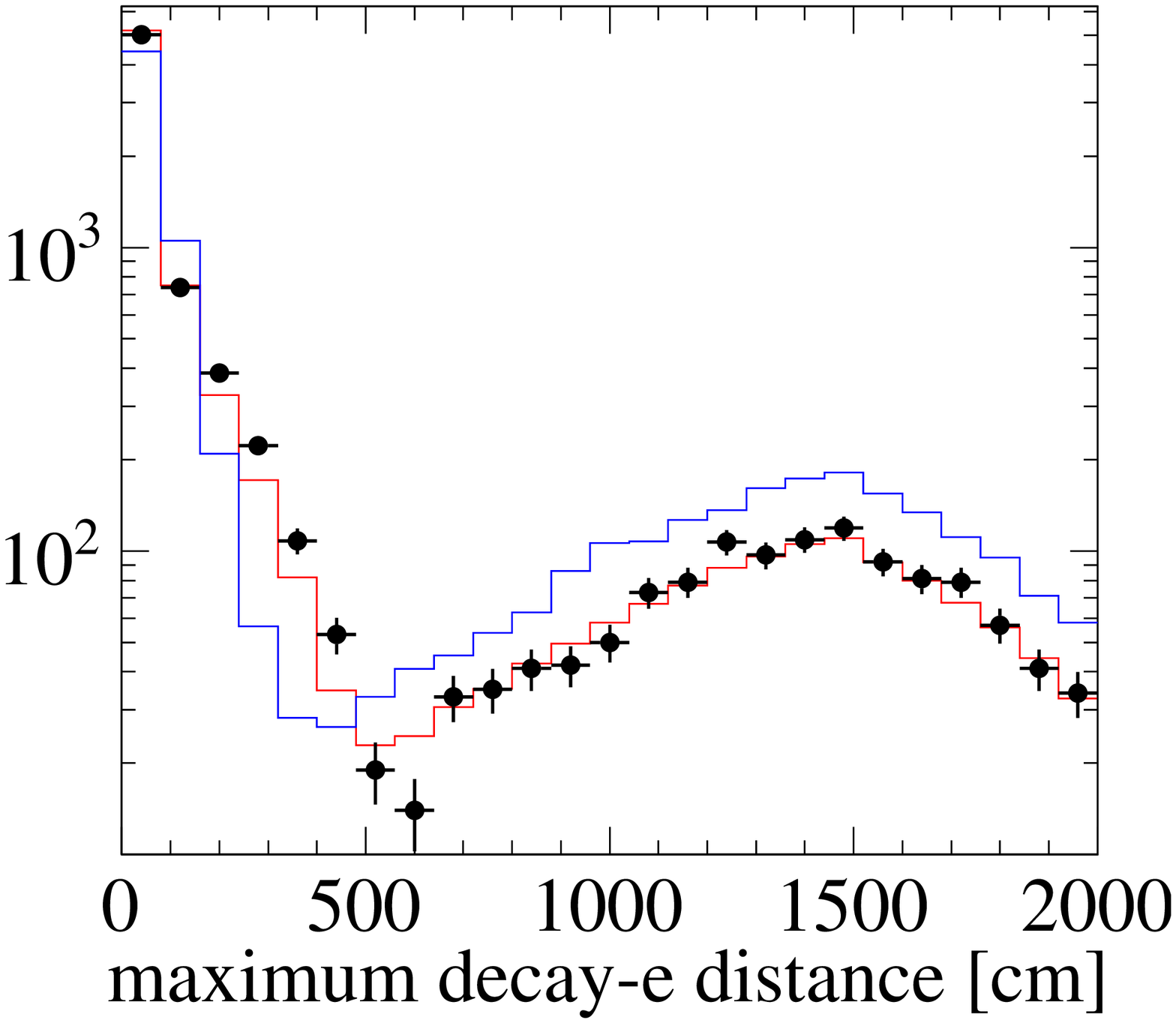}
		\caption{The 12 input variables to the multivariate analysis for signal (blue), background (red), and data (black), after precuts. Signal and background simulations are normalized to data.}
   \label{fig:mva_precut}
\end{figure*}

Finally, three variables related to particle identification are used: the number of e-like rings,
the number of decay electrons, and the maximum distance to any decay electron from the primary vertex.
Due to the large number of signal modes with one or more $\pi^{0}$s in the final state,
signal events are expected to have more e-like rings from their decays into photons.
Corresponding distributions for signal and background Monte Carlo (MC) after the pre-cuts are shown in Fig.~\ref{fig:mva_precut}.

These 12 variables are used in the construction of a multilayer perceptron (MLP)~\cite{Rosenblatt1963PRINCIPLESON}, which 
is trained on $n-\bar n$ signal and atmospheric neutrino background MC.
The MLP consists of a network of layers of nodes that are weighted and interconnected in order to optimize the discrimination 
between event types. 
Input variables form the input layer nodes and are combined in the MVA into a single node at the output layer,
which is the estimator describing how signal- or background-like an event is.
Between these layers there can be so-called hidden layers, whose structure and connectivity can be altered to optimize performance. 
In this analysis, a trial-and-error optimization for the hyper-parameters of the MLP structure was performed and the final 
structure was determined to be 1 hidden layer with $18$ hidden nodes.

\begin{figure} 
   \centering
  \includegraphics[width=0.90\columnwidth]{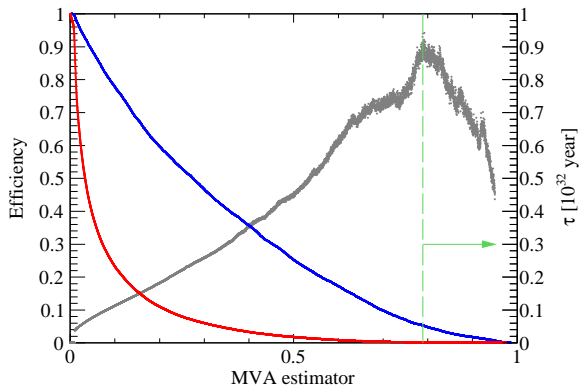}
		\caption{Signal (blue) and background (red) efficiency as a function of the MVA output estimator threshold. The expected sensitivity at each value of the estimator threshold as estimated using the Rolke method is shown in the gray curve.  The dashed line indicates the optimum cut point.}
   \label{fig:effC}
\end{figure}

The signal efficiency and background efficiency as a function of the estimator value is shown in Fig.~\ref{fig:effC}, where 
0 corresponds to background-like and 1 is signal-like.
A sensitivity analysis was performed assuming a 0.37 megaton$\cdot$years exposure and realistic systematic errors (described below) 
using the Rolke method~\cite{Rolke:2004mj} to determine the optimal cut position in the output estimator.
The optimized cut was found to be 0.789, where the signal (background) efficiency from the MVA alone is 5.0\% (0.1\%).
Combined with the pre-selection efficiency, the total signal efficiency is $4.1\%$
with an expected background of 0.56 events per year, or 9.3 events over the entire data period.
Selection efficiencies for each of the signal channels can be found in the last column of Table~\ref{tab:chnn} and Table~\ref{tab:chnp}.

\begin{figure} 
   \centering
      \includegraphics[width=0.95\linewidth]{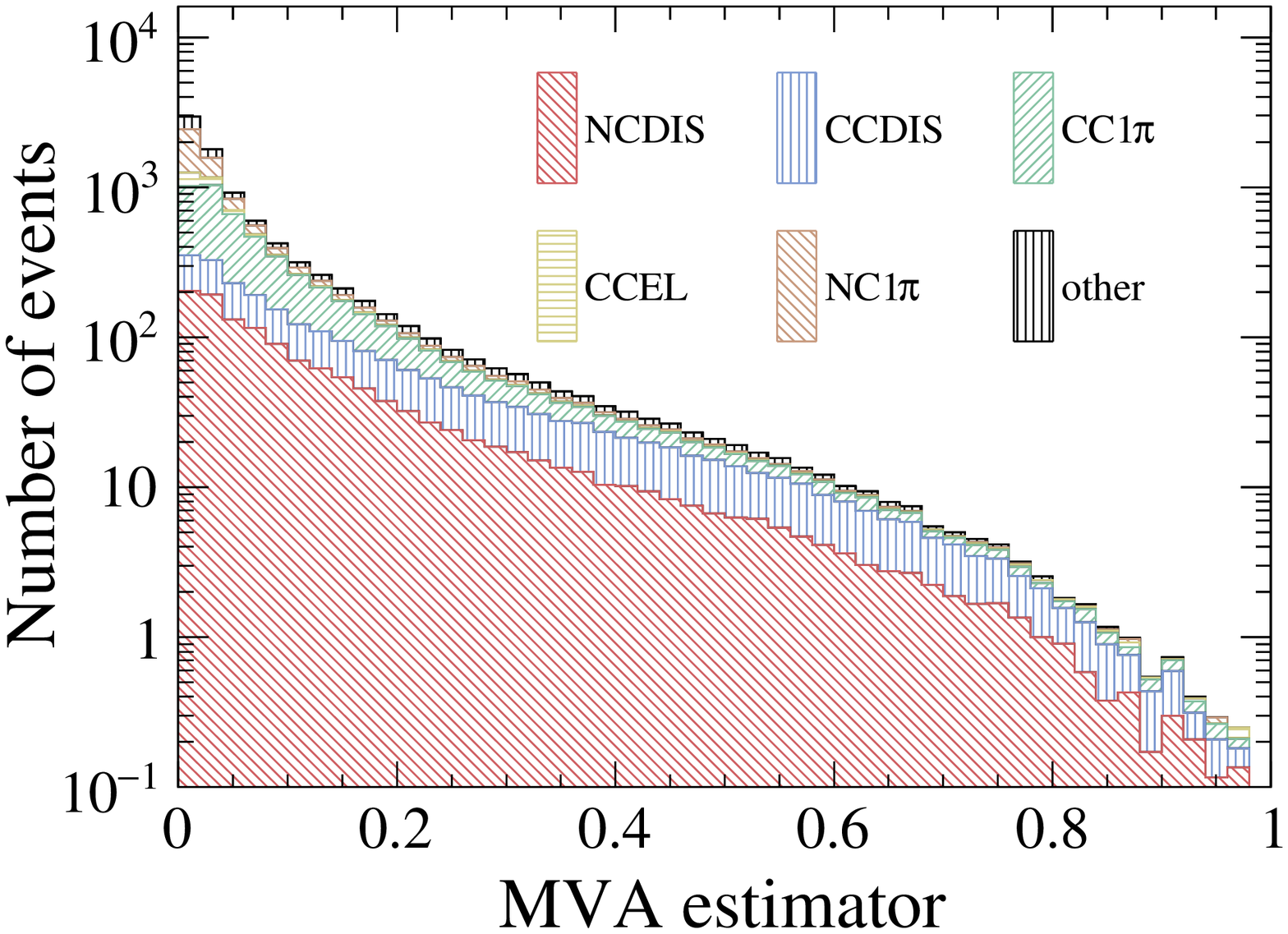}
		\caption{Remaining neutrino interaction backgrounds as a function of the MVA estimator after pre-selection and before MVA cut, broken-down into interaction channels, as shown in the legend.}
   \label{fig:remainingbkg}
\end{figure}

Among the multiple types of neutrino interactions, the dominant background in this analysis is from deep inelastic scattering (DIS), with secondary contributions from charged current pion production (CC 1$\pi$), neutral current pion production (NC 1$\pi$), and charged current elastic scattering (CC EL).
Figure~\ref{fig:remainingbkg} shows the remaining backgrounds before the MVA cut.
After applying the MVA cut, the remaining backgrounds in the final sample are shown in Table~\ref{tab:remainingbkg}, with $\nu_\mu+\bar\nu_\mu$ contributing 5.8 events, $\nu_e+\bar\nu_e$ contributing 3.5 events.
The contribution from $\nu_\tau+\bar\nu_\tau$ is less than $0.1$ event and is not shown in Table~\ref{tab:remainingbkg}.

\begin{table}
\caption{Summary of remaining atmospheric neutrino background events after the MVA selection and scaled to the full SK-I-IV livetime.}
   \label{tab:remainingbkg}
   \begin{tabular}{lccc}
      \hline\hline
		Channel & Events & $\nu_\mu+\bar\nu_\mu$ & $\nu_e+\bar\nu_e$ \\
      \hline
		NC DIS & 3.7 & - & - \\
		CC DIS & 3.6 & 2.0 & 1.6 \\
		CC 1$\pi$ & 1.1 & 0.7 & 0.4 \\
		CC EL  & 0.3 & 0.1 & 0.2 \\
		NC 1$\pi$ & 0.1 & - & - \\
		Other & 0.3 & - & - \\
      \hline
		Total & 9.3 & - & - \\
      \hline\hline
   \end{tabular}
\end{table}

\section{Systematic Estimation}
\label{sec:sys}

In this analysis systematic uncertainties are separated into two categories, those that arise from uncertainty in the physics modeling,
such as the hadronization process and final state interaction, and those related to the detector response and event reconstruction. 

\subsection{Modeling Uncertainties} 
\subsubsection{Signal}

Uncertainty in the momentum of the oxygen nucleons is expected to impact the resulting momentum of the $n-\bar n$ annihilation products.
A systematic uncertainty is derived from the difference between the default spectral function model (described in Sec.~\ref{sec:sim}) and the Fermi gas model~\cite{Smith:1972xh} used in the atmospheric neutrino simulation.
It yields an uncertainty in the signal efficiency of  7\%.

Measured uncertainties in the branching fraction of each annihilation channel also 
introduce a systematic uncertainty in the hadronization process, resulting in uncertainties in the pion multiplicity of signal events.
This uncertainty is accounted for by assigning uncertainties on the branching ratio of each channel listed in Table~\ref{tab:chnn} and Table~\ref{tab:chnp} based on the statistical uncertainty in the results from the Crystal Barrel~\cite{Klempt:2005pp, Amsler:2003bq} and the OBELIX experiments~\cite{Bressani:2003pv}.
They were then propagated to the analysis by reweighting the various final states accordingly and result in a 4\% uncertainty on the signal efficiency.

Final state interaction modeling is the dominant systematic error on the signal efficiency. 
To estimate this uncertainty, we generated separate MC sets, each with different FSI model parameters 
that control the strength of the interaction cross-sections and are allowed by fits to pion-nucleon scattering data~\cite{Jeffrey:2016}.
These MC samples were processed through the same event selection, and the largest change in the 
signal efficiency is taken as the uncertainty.
In this analysis, the largest deviation came from a variation with enhanced quasi-elastic scattering and absorption, 
but with decreased inelastic scattering, which produces fewer hadrons and thus lower efficiency.
The assigned uncertainty is 31\%.

\subsubsection{Background}

Uncertainties on the atmospheric neutrino background were calculated using the fit result from the SK atmospheric neutrino analysis~\cite{Abe:2017aap}.
A set of weights was constructed for each event, describing how it changes under a $1\sigma$ variation of each systematic error parameters used in that analysis. 
Applying these weights to the MC allows the uncertainty to be conservatively propagated to the background prediction.

The overall atmospheric neutrino flux normalization has an uncertainty of 15\%~\cite{Abe:2017aap} in the dominant background energy range between 1 and 10 GeV, 
resulting in a 7\% uncertainty in the background rate.
In total, the uncertainty introduced by modeling of the flux was estimated to be 8\%.
Neutrino PMNS oscillation parameter uncertainties, particularly from $\theta_{23}$, also introduce a 3\% uncertainty.
Uncertainties from the neutrino interaction modeling are the most significant contribution to the error budget.
The total uncertainty from neutrino interaction was estimated at 24\%, 
among which the main contribution was found to originate from uncertainties in the deep inelastic scattering model and its cross-section.

\subsection{Detector Systematics}

Uncertainties in the detector's energy scale and the reconstruction's ability to accurately identify the number of and particle type of each ring introduce uncertainties in both the signal efficiency and background rate. 
The energy scale uncertainty is evaluated using calibration sources and control samples, such as cosmic ray muons and their decay electrons~\cite{Abe:2017aap}, and is 3.3\% in SK-I, 2.8\% in SK-II, 2.4\% in SK-III, and 2.1\% in SK-IV. 
It results in a 5\% and 11\% uncertainty on the signal efficiency and background rate, respectively. 
Similarly, differences in the water quality in the top and bottom regions of the Super-K tank introduces an asymmetry in the energy scale 
that introduces an additional 4\% signal efficiency uncertainty and 6\% background rate uncertainty.

Ring counting introduces 2\% uncertainty in signal efficiency and 1\% in background rate.
This uncertainty is estimated by comparing the ring counting likelihood distribution of MC and a controlled sample data~\cite{Tanaka:2019}.

For MVA variables besides ring counting, energy scale, and non-uniformity, we use an inclusive controlled sample (FC data after precuts, before MVA), and compare data and MC prediction, as shown in Fig.~\ref{fig:mva_precut}.
The source uncertainties are assigned from the deviation of data and MC.
These source uncertainties are then propagated to efficiency uncertainties.

The individual systematic sources and their uncertainties are summarized in Table~\ref{tab:sys}, 
while the total efficiency and uncertainty are presented in Table~\ref{tab:ovesens}.

\begin{table}
   \caption{Summary of systematic uncertainties on the signal efficiency and backgrounds. 
     The atmospheric neutrino row represents the combined 
            uncertainty from modeling of their flux and interactions.}
   \label{tab:sys}
   \begin{tabular}{lrr}
      \hline\hline
      & Signal Efficiency & Background \\
      \hline
		Physics & & \\
      Hadronization & 4\% &-\\
      FSI & 31\% & -\\
      Fermi motion & 7\% & -\\
      Atmospheric $\nu$ & -& 24\%\\
      \hline
      Detector & & \\
      Energy scale & 5\% & 11\% \\
      Non-uniformity & 4\% & 6\% \\
      Ring counting & 2\% & 2\% \\
      Other MVA variables & 4\% & 7\% \\
      \hline
      Total & 33\% & 28\%\\
      \hline\hline
   \end{tabular}
\end{table}

\begin{table}
\caption{Overall efficiency and systematic uncertainty}
\label{tab:ovesens}
\begin{tabular}{lccc}
   \hline\hline
& Efficiency & Event rate & Systematics \\
        \hline
 			Signal & 4.1\%& - & 33\% \\
 			Background & - &0.56 / year & 28\% \\
          \hline\hline
       \end{tabular}
\end{table}

\section{Result}
\label{sec:result}

\begin{figure} 
   \centering
      \includegraphics[width=0.90\linewidth]{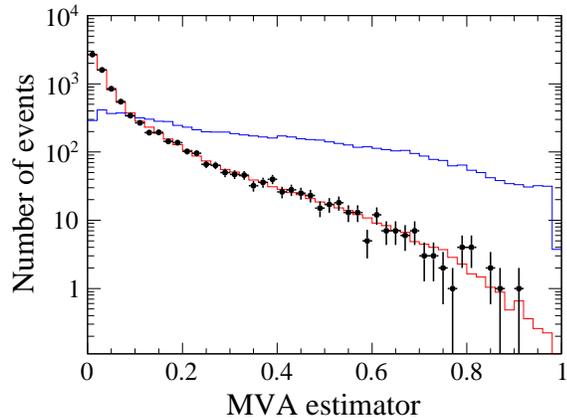}
		\caption{MVA output estimator after applying the analysis precut. The expected background is shown in red, the signal (scaled to data) in blue, and the data points are shown in black.}
   \label{fig:MVApre}
\end{figure}

\begin{figure} 
   \centering
      \includegraphics[width=0.90\linewidth]{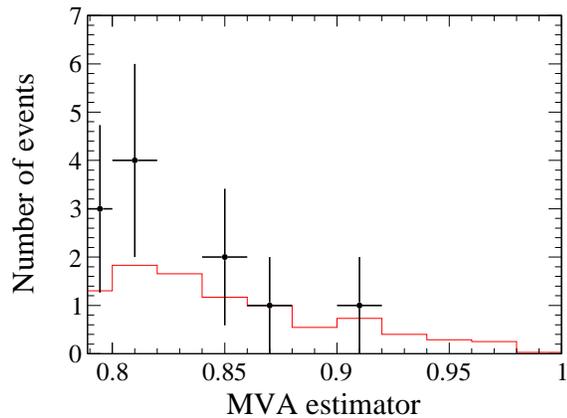}
		\caption{MVA output estimator after applying the cut at 0.789 in this variable. The expected background is shown in red, the signal in blue, and the data points are shown in black.}
   \label{fig:MVAMVA}
\end{figure}

This full SK-I-IV data set corresponds to an exposure of 0.37~megaton$\cdot$years. 
After applying the cuts above, $11$ events are found in data, which is consistent with the expected background of 
$9.3\pm 2.7$ events. 
Furthermore, data and MC are in good agreement both before (Fig.~\ref{fig:MVApre}) and after (Fig.~\ref{fig:MVAMVA}) the MVA cut. 
The input variables to the MVA show a similar agreement (Fig.~\ref{fig:mva_precut}).
Accordingly, we find no evidence for neutron-antineutron oscillations.


\begin{figure} 
   \centering
      \includegraphics[width=0.9\linewidth]{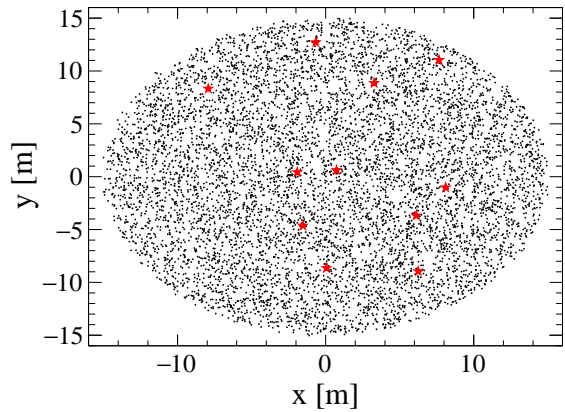}
      \includegraphics[width=0.9\linewidth]{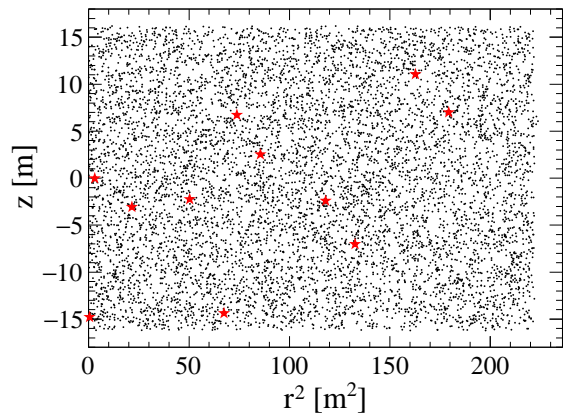}
		\caption{Spatial distribution of observed events after precuts (black), and after the MVA cut (red) in the plane perpendicular to the detector axis (top) and in the radial and axial direction (bottom).}
   \label{fig:space}
\end{figure}

\begin{figure} 
   \centering
      \includegraphics[width=0.90\linewidth]{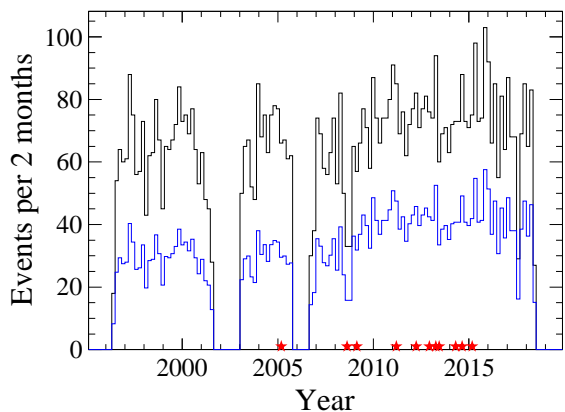}
		\caption{Distribution of observed events in time after precuts (black) and after the MVA cut (red). For comparison 
the background expectation after the MVA cut scaled by a factor of 500 is shown in blue.}
   \label{fig:date}
\end{figure}

Figure~\ref{fig:space} shows the 11 candidate events within the detector.
The spatial distribution is uniform as expected.
Figure~\ref{fig:date} shows the distribution in time. 
The dependence of events after precut on time is due to the live-time of SK.
Performing a K-S test on the distrbution yields a maximum distance between data 
and MC at 0.33.  
To determine the likelihood of this result, this procedure was repeated on simulated data sets with the
same size as the observation and assuming a constant rate.   
Among these pseudoexperiments, 14\% had a K-S distance larger than 0.33, indicating no significant 
deviation from the assumed uniform distribution and is consistent with the expectation 
from atmospheric backgrounds.

A comparison of the expected atmospheric neutrino background, signal efficiency and observed data in each SK run period 
is shown in Table.~\ref{tab:data1-4}.
Signal efficiencies and background rates are slightly different across these run periods, and the majority of candidate 
events are found in SK-IV, which has the longest livetime.
The Poisson probability of observing 9 or more events in SK-IV 
with an expectation of 5.5 events (ignoring systematic uncertainties) is $10.6\%$. 
This observation is similarly consistent with the background expectation.

\begin{table}
   \caption{Comparison of the expected atmospheric neutrino background, signal efficiency, livetime, and observation in each run period.}
   \label{tab:data1-4}
   \begin{tabular}{lrrrr}
      \hline\hline
      & SK-I & SK-II & SK-III & SK-IV\\
      \hline
      Efficiency     & 3.7\% & 3.3\% & 3.7\% & 4.4\% \\
      Background events & 1.98 & 1.03 & 0.74 & 5.50 \\
      Livedays       & 1489.2 & 798.6 & 518.1 & 3244.1 \\
      Candidates     & 0 & 1 & 1 & 9 \\
      \hline
   \end{tabular}
\end{table}


In the absence of a statistically significant excess in data, a lower limit is established.
To account for both statistical and systematic uncertainties, we used Rolke method in confidence interval calculation.
The observation limit on neutron lifetime is set at $3.6\times{10}^{32}$ years (90\% C.L.).
Equation~(\ref{eq:tranlim}) and $R = 0.517\times10^{23}$ s$^{-1}$~\cite{Friedman:2008es} are used to 
derive the corresponding limit on the $n-\bar n$ oscillation time, 
$\tau_{n\to\bar n} > 4.7\times10^{8}$ s.
A comparison between the expected sensitivity and this result is shown in Table.~\ref{tab:sens}.
Alternative calculations of the nuclear suppression factor $R$ can be found in Refs~\cite{Haidenbauer_2020, Oosterhof_2019}.

\begin{table}
   \caption{Expected and observed limits from the background-only hypothesis.}
   \label{tab:sens}
	   \begin{tabular}{lccc}
         \hline\hline
         &  Events &\  $T_{n-\bar n}$ ($10^{32}$ yrs) &\  $\tau_{n\to\bar n}$ ($10^8$ s)\\ 
         \hline
         Expected & 9.3 & 4.3 & 5.1 \\
         Observed & 11 & 3.6 & 4.7 \\
         \hline\hline
      \end{tabular}
\end{table}

Table~\ref{tab:comp} compares the present results with those from other bound neutron experiments 
and free neutron oscillation experiments.
Papers before the year 2000 typically report $\tau_{n\to \bar n}$ assuming $R=1\times10^{23}$/s, 
 and the previous SK result considered uncertainty in the theoretical prediction of $R$.
For better comparison and easier conversion, $\tau_{n\to \bar n}$ is presented as $\sqrt{T_{n-\bar n}/R}$ 
with corresponding nuclear suppression factor $R$ listed in Table~\ref{tab:comp}.
This analysis gives the most stringent limit on $n\to\bar n$ oscillation so far.

\begin{table*}
	\caption{Comparison of $n-\bar n$ oscillation searches from bound neutrons and free neutrons. All values of 
$\tau_{n\to \bar n}$ results are presented as $\sqrt{T_{n-\bar n}/R}$, where $R$ is the suppression factor used in each reference.}
	\label{tab:comp}
   \begin{tabular}{lrclc}
      \hline\hline
		& & $T_{n-\bar n} $($10^{32}$ years) & $R$ ($10^{23}$/s) & $\tau_{n\to \bar n}$($10^8$ s)\\
      \hline
		$^{16}$O & SK-I-IV (this study) & 3.6 & 0.517 & 4.7 \\
		$^{16}$O & SK-I~\cite{Abe:2011ky} (2015) & 1.9 & 0.517 & 3.4 \\
		$^{16}$O & Kamiokande~\cite{Takita:1986zm} (1986) & 0.4 & 0.517 & 1.6 \\
		$^2$H    & SNO~\cite{Aharmim:2017jna} (2017) & 0.1 & 0.25& 1.4 \\
		$^{56}$Fe& Soudan II~\cite{Chung:2002fx} (2002) & 0.7 & 1.4 & 1.3 \\
		$^{56}$Fe& Frejus~\cite{Berger:1989gw} (1990) & 0.7 & 1.4 & 1.2 \\
		$^{16}$O & IMB~\cite{Jones:1983ij} (1984) & 0.2 & 0.517 & 1.2 \\
		Free neutron & Grenoble~\cite{BaldoCeolin:1994jz} (1994) & - & - & 0.9 \\
      \hline\hline
   \end{tabular}
\end{table*}

\section{Conclusion}
\label{sec:conclusion}

We performed a $n-\bar n$ oscillation search with SK-I-IV data using a multi-variate analysis.
Compared to previous presults~\cite{Abe:2011ky}, the updated final state interaction model predicts fewer pions and less separation between signal and neutrino backgrounds.
With the advanced MVA method and the inclusion of multiple new variables, the sensitivity of this analysis is still greatly enhanced.

For the 0.37 megaton$\cdot$year exposure at SK, we observed $11$ events with an expected background of $9.3\pm 2.7$ events.
There is no statistically significant excess of data events, so 
a lower limit on the neutron lifetime is set at $3.6\times{10}^{32}$ years at 90\% C.L., corresponding to a lower limit on the neutron-antineutron oscillation time in $^{16}$O of $\tau_{n\to\bar n} >4.7\times10^{8}$ s.
This is the world's most stringent limit on neutron-antineutron oscillation so far, with 90\% improvement from the previous best limit~\cite{Abe:2011ky}.

\section*{Acknowledgments}
\input{SK-paper-acknowledgements-200529.tex}

\bibliography{apstemplate}

\end{document}

%% file: Authors-20200529.tex
\newcommand{\AFFicrr}{\affiliation{Kamioka Observatory, Institute for Cosmic Ray Research, University of Tokyo, Kamioka, Gifu 506-1205, Japan}}
\newcommand{\AFFkashiwa}{\affiliation{Research Center for Cosmic Neutrinos, Institute for Cosmic Ray Research, University of Tokyo, Kashiwa, Chiba 277-8582, Japan}}
\newcommand{\AFFipmu}{\affiliation{Kavli Institute for the Physics and
Mathematics of the Universe (WPI), The University of Tokyo Institutes for Advanced Study,
University of Tokyo, Kashiwa, Chiba 277-8583, Japan }}
\newcommand{\AFFmad}{\affiliation{Department of Theoretical Physics, University Autonoma Madrid, 28049 Madrid, Spain}}
\newcommand{\AFFubc}{\affiliation{Department of Physics and Astronomy, University of British Columbia, Vancouver, BC, V6T1Z4, Canada}}
\newcommand{\AFFbu}{\affiliation{Department of Physics, Boston University, Boston, MA 02215, USA}}
\newcommand{\AFFuci}{\affiliation{Department of Physics and Astronomy, University of California, Irvine, Irvine, CA 92697-4575, USA }}
\newcommand{\AFFcsu}{\affiliation{Department of Physics, California State University, Dominguez Hills, Carson, CA 90747, USA}}
\newcommand{\AFFcnm}{\affiliation{Institute for Universe and Elementary Particles, Chonnam National University, Gwangju 61186, Korea}}
\newcommand{\AFFduke}{\affiliation{Department of Physics, Duke University, Durham NC 27708, USA}}
\newcommand{\AFFfukuoka}{\affiliation{Junior College, Fukuoka Institute of Technology, Fukuoka, Fukuoka 811-0295, Japan}}
\newcommand{\AFFgifu}{\affiliation{Department of Physics, Gifu University, Gifu, Gifu 501-1193, Japan}}
\newcommand{\AFFgist}{\affiliation{GIST College, Gwangju Institute of Science and Technology, Gwangju 500-712, Korea}}
\newcommand{\AFFuh}{\affiliation{Department of Physics and Astronomy, University of Hawaii, Honolulu, HI 96822, USA}}
\newcommand{\AFFicl}{\affiliation{Department of Physics, Imperial College London , London, SW7 2AZ, United Kingdom }}
\newcommand{\AFFkek}{\affiliation{High Energy Accelerator Research Organization (KEK), Tsukuba, Ibaraki 305-0801, Japan }}
\newcommand{\AFFkobe}{\affiliation{Department of Physics, Kobe University, Kobe, Hyogo 657-8501, Japan}}
\newcommand{\AFFkyoto}{\affiliation{Department of Physics, Kyoto University, Kyoto, Kyoto 606-8502, Japan}}
\newcommand{\AFFliv}{\affiliation{Department of Physics, University of Liverpool, Liverpool, L69 7ZE, United Kingdom}}
\newcommand{\AFFmiyagi}{\affiliation{Department of Physics, Miyagi University of Education, Sendai, Miyagi 980-0845, Japan}}
\newcommand{\AFFnagoya}{\affiliation{Institute for Space-Earth Environmental Research, Nagoya University, Nagoya, Aichi 464-8602, Japan}}
\newcommand{\AFFkmi}{\affiliation{Kobayashi-Maskawa Institute for the Origin of Particles and the Universe, Nagoya University, Nagoya, Aichi 464-8602, Japan}}
\newcommand{\AFFpol}{\affiliation{National Centre For Nuclear Research, 02-093 Warsaw, Poland}}
\newcommand{\AFFsuny}{\affiliation{Department of Physics and Astronomy, State University of New York at Stony Brook, NY 11794-3800, USA}}
\newcommand{\AFFokayama}{\affiliation{Department of Physics, Okayama University, Okayama, Okayama 700-8530, Japan }}
\newcommand{\AFFosaka}{\affiliation{Department of Physics, Osaka University, Toyonaka, Osaka 560-0043, Japan}}
\newcommand{\AFFox}{\affiliation{Department of Physics, Oxford University, Oxford, OX1 3PU, United Kingdom}}
\newcommand{\AFFqmul}{\affiliation{School of Physics and Astronomy, Queen Mary University of London, London, E1 4NS, United Kingdom}}
\newcommand{\AFFregina}{\affiliation{Department of Physics, University of Regina, 3737 Wascana Parkway, Regina, SK, S4SOA2, Canada}}
\newcommand{\AFFseoul}{\affiliation{Department of Physics, Seoul National University, Seoul 151-742, Korea}}
\newcommand{\AFFsheff}{\affiliation{Department of Physics and Astronomy, University of Sheffield, S3 7RH, Sheffield, United Kingdom}}
\newcommand{\AFFshizuokasc}{\affiliation{Department of Informatics in
Social Welfare, Shizuoka University of Welfare, Yaizu, Shizuoka, 425-8611, Japan}}
\newcommand{\AFFstfc}{\affiliation{STFC, Rutherford Appleton Laboratory, Harwell Oxford, and Daresbury Laboratory, Warrington, OX11 0QX, United Kingdom}}
\newcommand{\AFFskk}{\affiliation{Department of Physics, Sungkyunkwan University, Suwon 440-746, Korea}}
\newcommand{\AFFtokyo}{\affiliation{The University of Tokyo, Bunkyo, Tokyo 113-0033, Japan }}
\newcommand{\AFFtodai}{\affiliation{Department of Physics, University of Tokyo, Bunkyo, Tokyo 113-0033, Japan }}
\newcommand{\AFFtit}{\affiliation{Department of Physics,Tokyo Institute of Technology, Meguro, Tokyo 152-8551, Japan }}
\newcommand{\AFFtus}{\affiliation{Department of Physics, Faculty of Science and Technology, Tokyo University of Science, Noda, Chiba 278-8510, Japan }}
\newcommand{\AFFtoronto}{\affiliation{Department of Physics, University of Toronto, ON, M5S 1A7, Canada }}
\newcommand{\AFFtriumf}{\affiliation{TRIUMF, 4004 Wesbrook Mall, Vancouver, BC, V6T2A3, Canada }}
\newcommand{\AFFtokai}{\affiliation{Department of Physics, Tokai University, Hiratsuka, Kanagawa 259-1292, Japan}}
\newcommand{\AFFtsinghua}{\affiliation{Department of Engineering Physics, Tsinghua University, Beijing, 100084, China}}
\newcommand{\AFFynu}{\affiliation{Department of Physics, Yokohama National University, Yokohama, Kanagawa, 240-8501, Japan}}
\newcommand{\AFFllr}{\affiliation{Ecole Polytechnique, IN2P3-CNRS, Laboratoire Leprince-Ringuet, F-91120 Palaiseau, France }}
\newcommand{\AFFbari}{\affiliation{ Dipartimento Interuniversitario di Fisica, INFN Sezione di Bari and Universit\`a e Politecnico di Bari, I-70125, Bari, Italy}}
\newcommand{\AFFnapoli}{\affiliation{Dipartimento di Fisica, INFN Sezione di Napoli and Universit\`a di Napoli, I-80126, Napoli, Italy}}
\newcommand{\AFFroma}{\affiliation{INFN Sezione di Roma and Universit\`a di Roma ``La Sapienza'', I-00185, Roma, Italy}}
\newcommand{\AFFpadova}{\affiliation{Dipartimento di Fisica, INFN Sezione di Padova and Universit\`a di Padova, I-35131, Padova, Italy}}
\newcommand{\AFFkeio}{\affiliation{Department of Physics, Keio University, Yokohama, Kanagawa, 223-8522, Japan}}
\newcommand{\AFFwinnipeg}{\affiliation{Department of Physics, University of Winnipeg, MB R3J 3L8, Canada }}
\newcommand{\AFFkcl}{\affiliation{Department of Physics, King's College London, London, WC2R 2LS, UK }}
\newcommand{\AFFwarwick}{\affiliation{Department of Physics, University of Warwick, Coventry, CV4 7AL, UK }}
\newcommand{\AFFral}{\affiliation{Rutherford Appleton Laboratory, Harwell, Oxford, OX11 0QX, UK }}
\newcommand{\AFFwu}{\affiliation{Faculty of Physics, University of Warsaw, Warsaw, 02-093, Poland }}
\newcommand{\AFFbcit}{\affiliation{Department of Physics, British Columbia Institute of Technology, Burnaby, BC, V5G 3H2, Canada }}

\AFFicrr
\AFFkashiwa
\AFFmad
\AFFbu
\AFFuci
\AFFcsu
\AFFcnm
\AFFduke
\AFFllr
\AFFfukuoka
\AFFgifu
\AFFgist
\AFFuh
\AFFicl
\AFFbari
\AFFnapoli
\AFFpadova
\AFFroma
\AFFkcl
\AFFkeio
\AFFkek
\AFFkobe
\AFFkyoto
\AFFliv
\AFFmiyagi
\AFFnagoya
\AFFkmi
\AFFpol
\AFFsuny
\AFFokayama
\AFFosaka
\AFFox
\AFFral
\AFFseoul
\AFFsheff
\AFFshizuokasc
\AFFstfc
\AFFskk
\AFFtokai
\AFFtokyo
\AFFtodai
\AFFipmu
\AFFtit
\AFFtus
\AFFtoronto
\AFFtriumf
\AFFtsinghua
\AFFwarwick
\AFFwinnipeg
\AFFynu
\AFFwu
\AFFbcit

\author{K.~Abe}
\AFFicrr
\AFFipmu
\author{C.~Bronner}
\AFFicrr
\author{Y.~Hayato}
\AFFicrr
\AFFipmu
\author{M.~Ikeda}
\author{S.~Imaizumi}
\AFFicrr
\author{H.~Ito}
\AFFicrr 
\author{J.~Kameda}
\AFFicrr
\AFFipmu
\author{Y.~Kataoka}
\AFFicrr
\author{M.~Miura} 
\author{S.~Moriyama} 
\AFFicrr
\AFFipmu
% ...


\collaboration{The Super-Kamiokande Collaboration}
\noaffiliation

%% file: SK-paper-acknowledgements-200529.tex

We gratefully acknowledge the cooperation of the Kamioka Mining and Smelting Company.
The Super-Kamiokande experiment has been built and operated from funding by the 
Japanese Ministry of Education, Culture, Sports, Science and Technology, the U.S.
Department of Energy, and the U.S. National Science Foundation. Some of us have been 
supported by funds from the National Research Foundation of Korea NRF‐2009‐0083526
(KNRC) funded by the Ministry of Science, ICT, and Future Planning and the Ministry of
Education (2018R1D1A3B07050696, 2018R1D1A1B07049158), 
the Japan Society for the Promotion of Science, the National
Natural Science Foundation of China under Grants No. 11620101004, the Spanish Ministry of Science, 
Universities and Innovation (grant PGC2018-099388-B-I00), the Natural Sciences and 
Engineering Research Council (NSERC) of Canada, the Scinet and Westgrid consortia of
Compute Canada, the National Science Centre, Poland (2015/18/E/ST2/00758),
the Science and Technology Facilities Council (STFC) and GridPPP, UK, the European Union's 
Horizon 2020 Research and Innovation Programme under the Marie Sklodowska-Curie grant
agreement no.754496, H2020-MSCA-RISE-2018 JENNIFER2 grant agreement no.822070, and 
H2020-MSCA-RISE-2019 SK2HK grant agreement no. 872549.